\documentclass[superscriptaddress,amsmath,twocolumn,amssymb,longbibliography,10pt]{revtex4-2}

\usepackage[utf8]{inputenc}
\usepackage[T1]{fontenc}
\usepackage{lmodern}
\usepackage{amsthm}
\newtheorem{theorem}{Theorem}
\usepackage{subcaption}
\usepackage{caption}
\usepackage{stackengine}
\usepackage{amsmath}
\usepackage{amssymb}
\usepackage{enumitem}
\usepackage{graphicx}
\usepackage{braket}
\usepackage{bm}
\usepackage{algorithm}
\usepackage{algpseudocode}
\usepackage{float}
\usepackage{xcolor}
\usepackage[colorlinks=true,citecolor=blue,urlcolor=blue, linkcolor=blue]{hyperref}

\begin{document}

\title{Efficient quantum Gibbs sampling of stabilizer codes using hybrid computation}

\author{Ivan H.C. Shum}
\email{hcs64@cam.ac.uk}
\affiliation{Department of Applied Mathematics and Theoretical Physics, University of Cambridge}

\author{Angela Capel}
\email{ac2722@cam.ac.uk}
\affiliation{Department of Applied Mathematics and Theoretical Physics, University of Cambridge}
    
\begin{abstract}

We present hybrid Gibbs sampling algorithms for the stabilizer code Hamiltonians of the rotated surface code and the toric code with only local quantum algorithms, using $\sim L/2$ quantum circuit depth to prepare the Gibbs state of the rotated surface code Hamiltonian, and $L$ quantum circuit depth to prepare the Gibbs state of the toric code Hamiltonian, being $L$ the side of the side of the square lattice. We further show that if we allow for non-local gates, the Gibbs state of the periodic 1D Ising model can be prepared in logarithmic depth and linearly many simultaneous measurements. 

\textcolor{red}{}

\end{abstract}

\maketitle

\tableofcontents

\section{Introduction}

The Gibbs state of a system, given by $\rho=\dfrac{e^{-\beta H}}{Z(\beta)}$ where $H$ is the Hamiltonian and $\beta = 1/k_B T$ is the inverse temperature, is known to encode the equilibrium properties of the system. Sampling from such a state for classical statistical mechanics models is usually done by Markov Chain Monte Carlo methods such as Metropolis-Hastings \cite{Levin.2008},  which are typically efficient at high temperatures \cite{Martinelli.1999} and expected to be efficient in more generality \cite{Brooks.2011}. The quantum equivalent of such algorithms has been first found in \cite{QMH2009}, setting up the Markov chain only transitioning over computational basis states \cite{Yung2010}, and has been subsequently extended in \cite{Rall.2023,Wocjan.2023,Jiang.2024,Gilyen.2024}, with provable guarantees only under further theoretical assumptions. 

Recent advances in approximations of the quantum Gibbs state have generally been in one of two directions, either via dissipation, by simulating the Lindbladian evolution of the system, as discussed in \cite{Rouze2024}, which induces a quantum Markov semigroup with the Gibbs state as the stationary state and in some cases can be shown to mix in time scaling polynomially with the system size \cite{Kastoryano2016GibbsSamplersCommuting,Ding2024EfficientQuantumGibbs,Gilyen.2024,Bardet.2023,Bardet.2024,Capel2024Gibbs,Capel2024RapidThermalizationDissipative,Bakshi.2024,Rouze2024}; or by estimating the Boltzmann weights by a polynomial and reducing the problem for commuting $k$-local Hamiltonians to a classical decoding problem \cite{schmidhuber2025hamiltoniandecodedquantuminterferometry}, achieving a polynomial depth preparation for Pauli Hamiltonians with a symplectic code of constant dimension. Recent advances in exact preparations of the Gibbs state map commuting Pauli Hamiltonians to classical statistical models, for which sampling methods are well known to sample from the exact distribution \cite{Gibbs2025,hwang2024gibbsstatepreparationcommuting,schmidhuber2025hamiltoniandecodedquantuminterferometry}. 

The algorithms discussed in our paper build upon the approach derived in \cite{Gibbs2025} and prepare the exact Gibbs distribution for two stabilizer codes, the \textbf{rotated surface code} defined on an open $(L+1)\times(L+1)$ lattice and the \textbf{2D toric code} defined on a $L\times L$ torus, by first preparing a simpler distribution and then encoding the excited states using a low depth gateset of controlled-NOT($CX$) and Hadamard($H$) gates. We improve on the depth of $\mathcal O(L^3)$ obtained in \cite{Gibbs2025} for the toric code, and present $\mathcal O(L)$ depth circuits to encode the excited states given a sample of the syndrome that is in principle unknown, which saturates the Lieb-Robinson bound for preparing states exhibiting quantum topological order \cite{Bravyi2006}. For the rotated surface code, which is a slight generalization of the defected toric code studied in \cite{hwang2024gibbsstatepreparationcommuting}, we improve with an $\mathcal O(L)$ depth circuit upon the best previous result of $\mathcal O(L^2)$, to the best of our knowledge. We have made use of the fact that the 2 codes are unitarily mapped to 1D Ising models with a low number of gates, first shown in \cite{Gibbs2025}, and therefore it is possible to carry out the sampling process classically. We also consider an algorithm that uses quantum measurements, in particular using the fact that in principle we can generate statistics for $n$ independent Bernoulli variables with a non-Clifford circuit of depth 1 combined with a simultaneous measurement over all qubits. This will result in an algorithm that Gibbs samples the 1D Ising chain in logarithmic depth and a simultaneous measurement of all relevant qubits. 

The first part of the paper will discuss efficient algorithms to do so, by reversing the process: starting from the stabilizer Hamiltonian $H_{code}$ and decoupling it into two subsystems that act on disjoint qubits non-trivially and are of purely composed of $X$ and $Z$ operators only. Hence, when mentioning our progress when going through the algorithm, we will start from the stabilizer Hamiltonian and evolve through a quantum circuit, mentioning when we have arrived at two decoupled systems, after which all $CX$ gates should be interpreted as classical $XOR$, and $q_x$ will be the sites with $X$ interactions on them. We will also discuss how to perform the classical Gibbs sampling in the appendix for completeness. 

\begin{theorem}
There is an $\mathcal O(L)$ depth local mapping from the both the 2D toric code and the 2D rotated surface code, to two decoupled systems that can be identified as 1D Ising models. 
\end{theorem}

We show that we can reduce the sampling process to a classical sampling process for both systems, in the same sense as in \cite[Theorem 2]{schmidhuber2025hamiltoniandecodedquantuminterferometry} to obtain the syndrome, with further classical $XOR$ manipulations, a layer of quantum $H$ (Hadamard) gates, followed by $\mathcal O(L)$ layers of $CX$ gates, where $L$ is the size of the side of the system (and thus of order $N^{1/2}$, with $N$ the total number of qubits), and are inherently quantum as they act on superposition of computational basis states, creating entanglement. The sampling process can be performed classically due to identifying the surface code Hamiltonians with being unitarily equivalent to decoupled classical Ising models, one for the $X$ subsystem and one for the $Z$ subsystem. Our main result is that we only need half the depth when processing the classical sampling result using a quantum local circuit, compared to current lowest depth local circuits that encode a logical state \cite{Claes2025, Shum2025}. The reduction in depth is due to deferring most of the state preparation before entanglement to classical pre-processing, staying in the computational basis and hence being perfectly possible to simulate efficiently classically. 

The second part of the paper discusses a quantum sampling algorithm, with stochasticity coming from measuring uncorrelated qubits rotated to match the Boltzmann weights of any real temperature $\beta$. 

\begin{theorem}
There is a local algorithm that samples the Gibbs  state of the 2D toric code at any positive temperature with a pre-processing $\mathcal O(L)$ depth circuit, $2L^2+ \mathcal O(1)$ measurements in the computational basis, of which $2L^2-2$ can be performed simultaneously, followed by a post-processing $\mathcal O(L)$ depth circuit. 
\end{theorem}

We will also see how the same idea applies to Gibbs sampling the $n$ qubit Ising loop with logarithmic depth circuits and $n+\mathcal O(1)$ measurements of which $n+1$ can be performed simultaneously. This can be considered as a reason to why we enforce locality in the previous attempt to Gibbs sample the stabilizer codes; it is well known in realizations of preparations of the encoding of logical states of the toric code that using local gates is simpler to realize, as is the choice made in \cite{Realization2021}. We note that this algorithm is hard to generalize; it works for an Ising loop type non-homogeneous Hamiltonian, but it is difficult to generalize to say, the 1D transverse field Ising model, where the perturbing transverse field also changes the eigenvectors, and hence although the Boltzmann weights are easy to prepare using the exact solution, to entangle qubits to the correct eigenstates remains a problem. It is also difficult to generalize to higher dimensional Ising loops, say defined over $\mathbb{Z}_n^d$, since the number of non-independent terms scales with $n^d$ and hence we require more classical resources scaling with $n^d$ to fix eigenvalues of the Hamiltonian. 

\begin{theorem}
There is a non-local algorithm that Gibbs samples the $n$ qubit 1D Ising model with periodic boundary conditions with a pre-processing $\mathcal O(1)$ depth circuit, $n+1$ measurements in the computational basis of which $n-1$ can be performed simultaneously, followed by a post-processing $\mathcal O(\log n)$ depth circuit. 
\end{theorem}

\section{Notation, convention, explanation of algorithms}

Following a recent previous paper in a similar direction \cite{Gibbs2025}, we will work in the Heisenberg picture, where we evolve the stabilizers composing each Hamiltonian by $CX$ gate conjugation. The idea is to evolve the Hamiltonian to a state where the computational basis is the eigenbasis, and therefore many of the computations used will be completely classical. 

\begin{figure}[h!] 
    \centering
    \includegraphics[width=8cm]{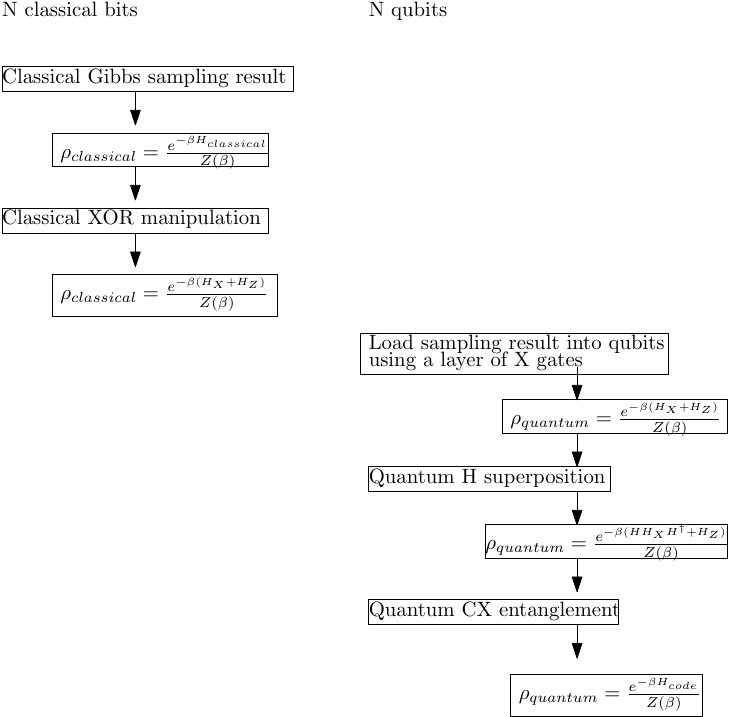}
    \caption{Flow of algorithm}
    \label{fig:flow}
\end{figure}

Fig. \ref{fig:flow} captures the idea of our algorithm to Gibbs sample stabilizer Hamiltonians $H_{code}$. It remains to see if we can write $UH_{code}U^\dagger=H^{\dagger}H_xH+H_z$, where $U$ is a unitary made up of layers of $CX$ gates, $H_x, H_z$ are sums of $Z$ operators acting on disjoint sets of qubits $q_x, q_z$, that are easy to sample from in the corresponding classical case; and $H=\bigotimes_{i\in q_x}H_i$ is a tensor product of Hadamard gates on a selected subset of the qubits, converting all terms in $H_x$ to purely $X$ operators. 

We will use an arrow to represent $CX$ gates, with the tail marking the control qubit and the arrowhead pointing at the target qubit. When with no further specification, $CX$ denotes the gate with control on first qubit and target on the second. The convention for the classical $XOR$ is the same, i.e. $XOR(a,b)=(a, a\oplus b)$. 

Lone $X$ terms will be marked by a red cross, while lone $Z$ terms will be marked by a blue circle. Due to the nature of $CX$ gates, $X$ interactions are mapped to $X$ interactions and $Z$ interactions are mapped to $Z$ interactions. Hence we will use a chain of red/blue segments to represent the $X$/$Z$ interactions respectively. 

\begin{figure}[h!] 
    \centering
    \includegraphics[width=3cm]{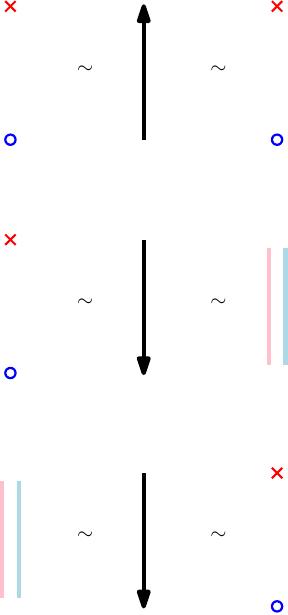}
    \caption{Basic operations after conjugation of $CX$ gates. We denote the sum of a real linear combination of stabilizers by red/blue cross/circles, lines and polygons, with red corresponding to purely $X$ interactions and blue corresponding to purely $Z$ interactions. An arrow is used to signify a $CX$ gate conjugation, pointly from the source to the target. }
    \label{fig1}
\end{figure}

Denoting the Pauli matrices by $\{I, X, Y, Z\}$ and noting $CX^\dagger=CX, CX^2=I \otimes I$, we remind the reader of the following results: 
\begin{align*}
    CX(I\otimes X)CX=I \otimes X\, ,& \quad CX(X \otimes X)CX=X \otimes I \, ,\\
CX(X \otimes I)CX=X \otimes X \, , & \quad  CX(Z \otimes I)CX=Z \otimes I \, ,\\
 CX(Z \otimes Z)CX=I \otimes Z \, , & \quad  CX(I \otimes Z)CX=Z \otimes Z \, ,
\end{align*}
which will be used extensively in the algorithms that follows. In Fig. \ref{fig1}, we present a diagrammatic notation that may be helpful for understanding the construction of our algorithms. 

All of the circuits presented will be in layers of $CX$ gates that mutually commute: recall that a sufficient criterion for so is the intersection of control qubits and target qubits being empty. 

The algorithms are based on two observations: the first is that the stabilizer codes above can be shown to be unitarily equivalent (with only local gates) to classical statistical mechanics models. For the rotated surface code, it is equivalent to a uniform magnetic field 
\begin{equation*}
    H_{UMF}=-\sum_{i\in \{Q\backslash q\}} Z_i \, ,
\end{equation*}
where $Q$ is the set of qubits and $q$ is an arbitrary qubit in $Q$. It is clear why $q$ can be chosen to be arbitrary: the $SWAP(a,b)$ gate can be implemented by $CX(a, b)CX(b, a)CX(a, b)$, where $a, b$ label qubits. For the $L\times L$ toric code, it is equivalent to 2 decoupled Ising loops of length $L$: 
\begin{equation*}
    H_{Loops}=-\hspace{-0.2cm}\sum_{i=1}^{L^2-1} \hspace{-0.1cm}Z_i Z_{i+1}\hspace{-0.1cm}-\hspace{-0.1cm}Z_{L^2}Z_1\hspace{-0.1cm}-\hspace{-0.2cm}\sum_{i=L^2+1}^{2L^2-1}  \hspace{-0.1cm}Z_i Z_{i+1}\hspace{-0.1cm}-\hspace{-0.1cm}Z_{2L^2}Z_{L^2+1}  ,
\end{equation*}
where again the labelling of the $2L^2$ qubits can be arbitrary due to the $SWAP$ gate argument above. However, we will note that for optimal depth the labelling of qubits is important only because we enforce locality. 

The second observation is that the quantum $CX$ is entirely identical to the classical $CNOT$ if the input is known to be a computational basis state. Thus we can prepare the Gibbs state of the classical statistical mechanics model, then process the state classically, and encode the qubits into this computational basis state; before applying a layer of $H$ gates, after which we necessarily have to use quantum $CX$ gates to get to the Gibbs state. This idea is useful only because classical computation is known to be more accurate and efficient with today's technological standards. Hence, working in the Heisenberg picture and starting from the stabilizer Hamiltonian, it suffices to decouple the $X$ subsystem from the $Z$ subsystem using $CX$ gates, and apply $H$ gates onto all of the sites hosting an $X$ operator, to obtain a Hamiltonian that is diagonal in the computational basis: Gibbs sampling from this Hamiltonian is then entirely classical. 

We then note that a similar idea can be used to prepare ground states of both stabilizer codes; since unitary gates map ground states to ground states, we only need to prepare a ground state of the decoupled Hamiltonian, which is trivial: on sites with $Z/X$ terms, initialise to $\ket{0}/\ket{+}$ respectively. Therefore one can expect that preparing the ground state of the stabilizer codes may be simpler than encoding logical information stored in physical qubits, a distinction first made in \cite{Chen2024}. 

The result is using a quantum circuit of depth $L/2$ to prepare the Gibbs state of the rotated surface code, compared to the quickest local encoding algorithm we know of in \cite{Claes2025} with a local circuit of depth $L$; and a quantum circuit of depth $L$ to prepare the Gibbs state of the toric code, compared to the quickest local encoding algorithm we know of in \cite{Shum2025}. 

\section{Hybrid Gibbs sampling the rotated surface code}

The relevant code for this section and the next are attached at \href{https://github.com/ihcs64/gibbsStatePreparation}{https://github.com/ihcs64/gibbsStatePreparation}, using the Stim library \cite{Gidney2021}. The symbol $f(x)\sim g(x)$ denotes $\lim_{x\rightarrow\infty} f(x)/g(x)=1$. 

We consider the rotated surface code defined on the $(L+1)^2$ lattice points of $[0, L]\times [0,L]$. Each square of the lattice is identified by the coordinates of its centre, with the convention that the lower left square (with coordinates $(1/2, 1/2)$) has an $X^{\otimes 4}$ operator. 

The rotated surface code is then given by the Hamiltonian 
\begin{align*}
    H_{RSC} =& -\Big(\sum\limits_{\{i,j,k,l\}\in  V_{odd}+}X_i X_j X_k X_l +\sum_{\{i,j\}\in  \partial V_{vert}}X_i X_j\Big) \\
    & -\Big(\sum_{\{i,j,k,l\}\in  V_{even}}Z_i Z_j Z_k Z_l\sum_{\{k,l\}\in  \partial V_{hor}}Z_k Z_l \Big) \, ,
\end{align*}
where $V_{even/odd}$ is the set of squares, labelled with coordinates $(x, y)$ such that $x+y$ is even/odd respectively, and $\partial V_{hor/vert}$ are the segments along the 2 horizontal/vertical boundaries, that share an edge of an even/odd square respectively. 

In particular, we remind the reader of the convention we have adopted, with $X$ boundary conditions on left/right, and $Z$ boundary conditions on top/bottom. Each square is labelled with coordinates increasing to the right and upwards and identified with the coordinates of the lower left vertex, with the standard convention that the lower left square (with coordinates $(0, 0)$) has an $X^{\otimes 4}$ operator. 

\begin{figure}[h!] 
    \centering
    \includegraphics[width=8cm]{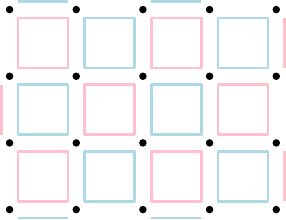}
    \caption{$4\times 3$ rotated surface code; the even sites host pink/$X$ operators, while the odd sites host blue/$Z$ operators}
    \label{fig:RSC}
\end{figure}

For example, for a $4\times 3$ grid, the initial Hamiltonian is represented diagrammatically in Fig. \ref{fig:RSC}.

To start describing our algorithm, we focus on $L$ even, then extend by a circuit of depth $\mathcal O(1)$ to the case where $L$ is odd. It is natural to consider one of the logical operators together with the Hamiltonian $H_{RSC}$, such that the $(L+1)^2$ operators generate the Pauli group. We will choose the representatives of a vertical line of $Z$ along the line $x=L/2$, and a vertical line of $X$ along the line $y=L/2$. 

\begin{figure}[h!] 
    \centering
    \includegraphics[width=8cm]{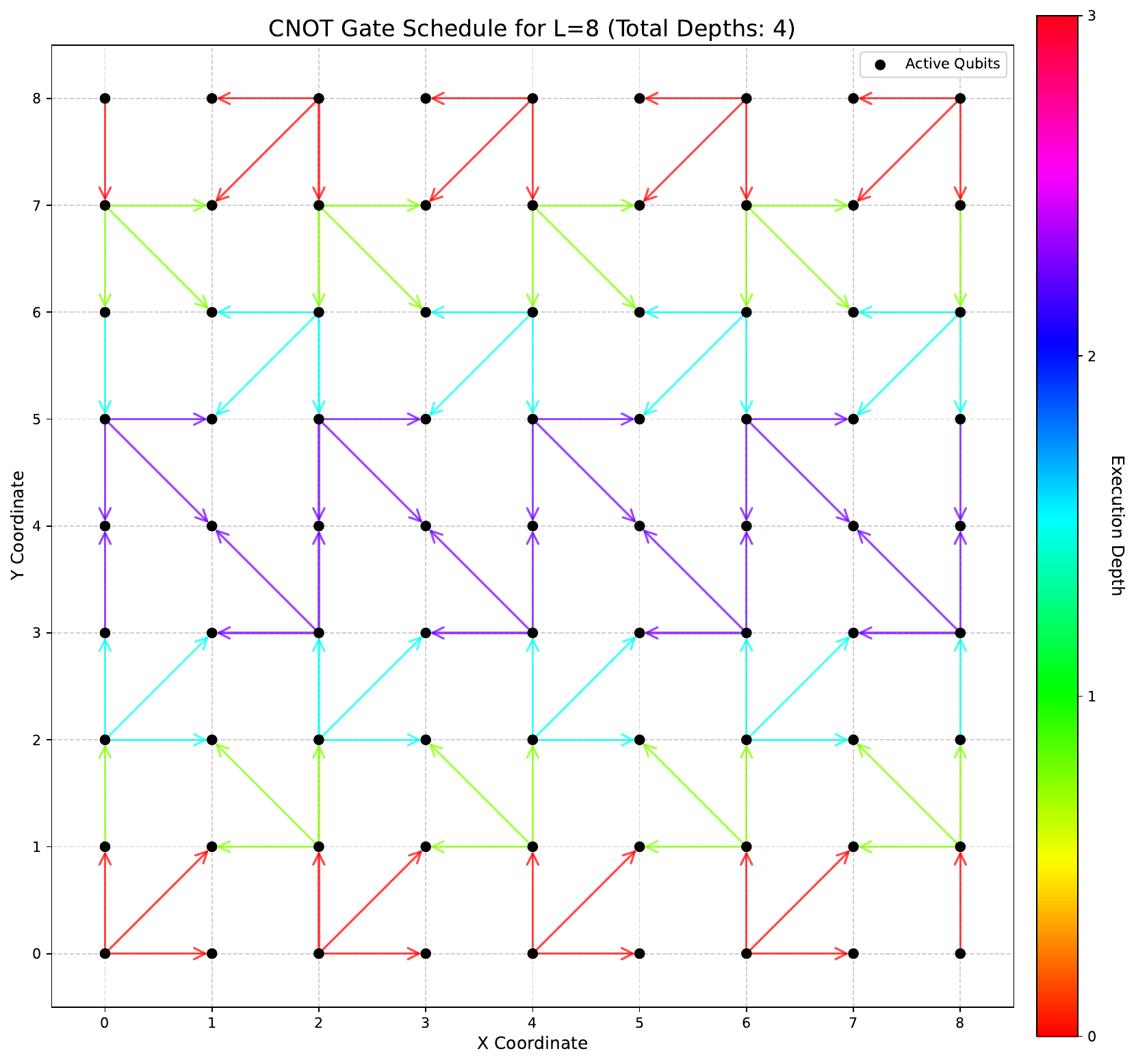}
    \caption{First step of algorithm when $L=6$}
    \label{fig:algoRSC}
\end{figure}

The $CX$ gates needed to evolve $H_{RSC}$ into 2 decoupled systems are then described in Fig. \ref{fig:algoRSC}. The $i^{th}$ step of the algorithm reduces all of the $X$ operators of the $i^{th}$ row, counting from bottom and from top, to single site operators. 

\begin{figure}[h!] 
    \centering
    \includegraphics[width=8cm]{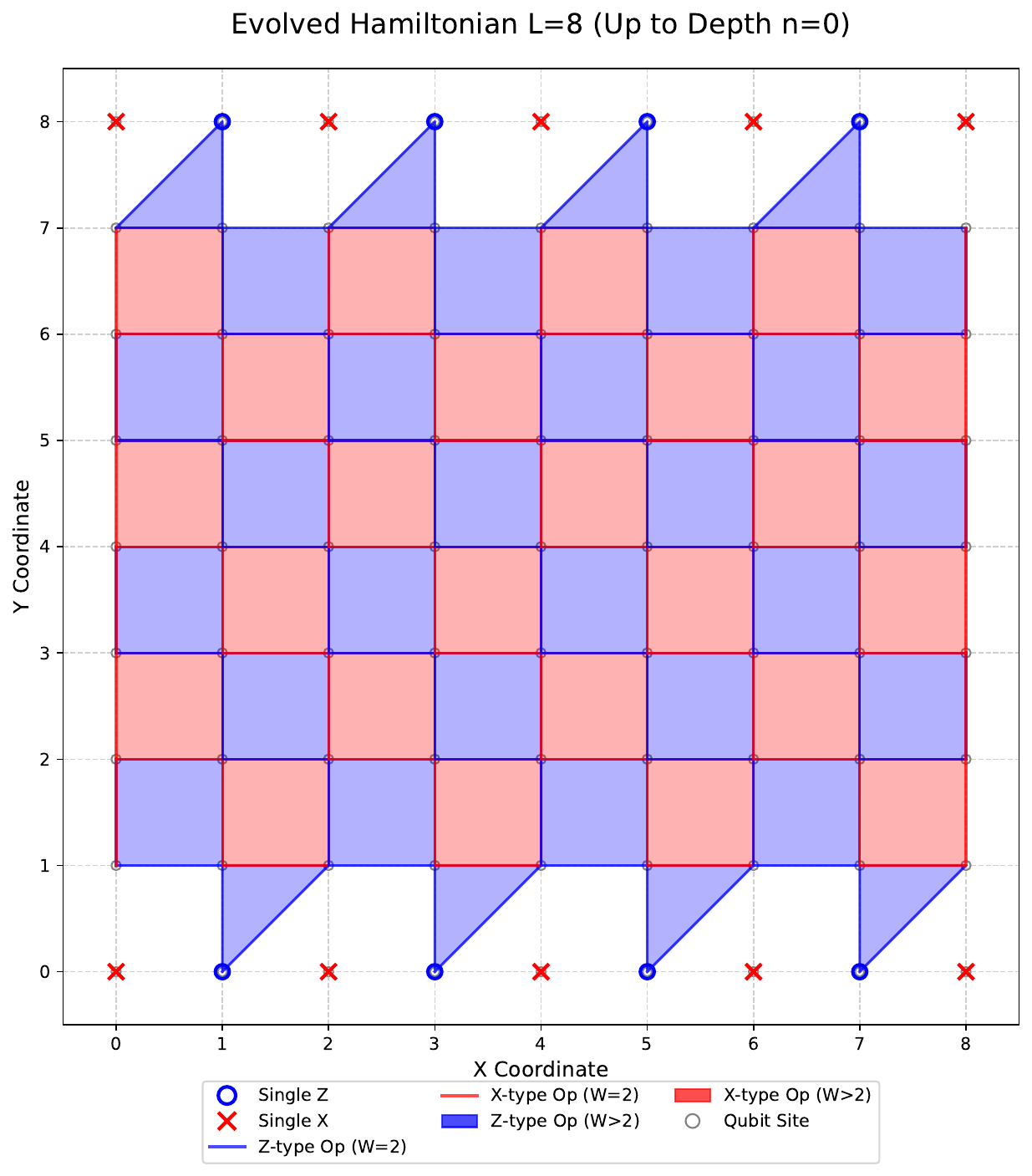}
    \caption{Resulting Hamiltonian after the first step of quantum circuit evolution}
    \label{fig:Ham1ststepRSC}
\end{figure}

We describe the choice taken in the bottom row: reducing each operator to its lower left site; and the description for all other rows follow, by noting that the rotated surface code is invariant under a rotation by $\pi$ and hence we can apply the same idea, rotated, to the top row; in particular, we note that we have stabilised all operators not in the top and bottom row, the $Z$ operators in the top and bottom rows are reduced to three sites, two of which form the boundary condition of the $L\times (L-2)$ rotated surface code, and the $Z$ boundary conditions are all reduced to single site as well, in an alternating pattern. The Hamiltonian obtained after this step is represented in Fig. \ref{fig:Ham1ststepRSC}.

\begin{figure}[h!] 
    \centering
    \includegraphics[width=8cm]{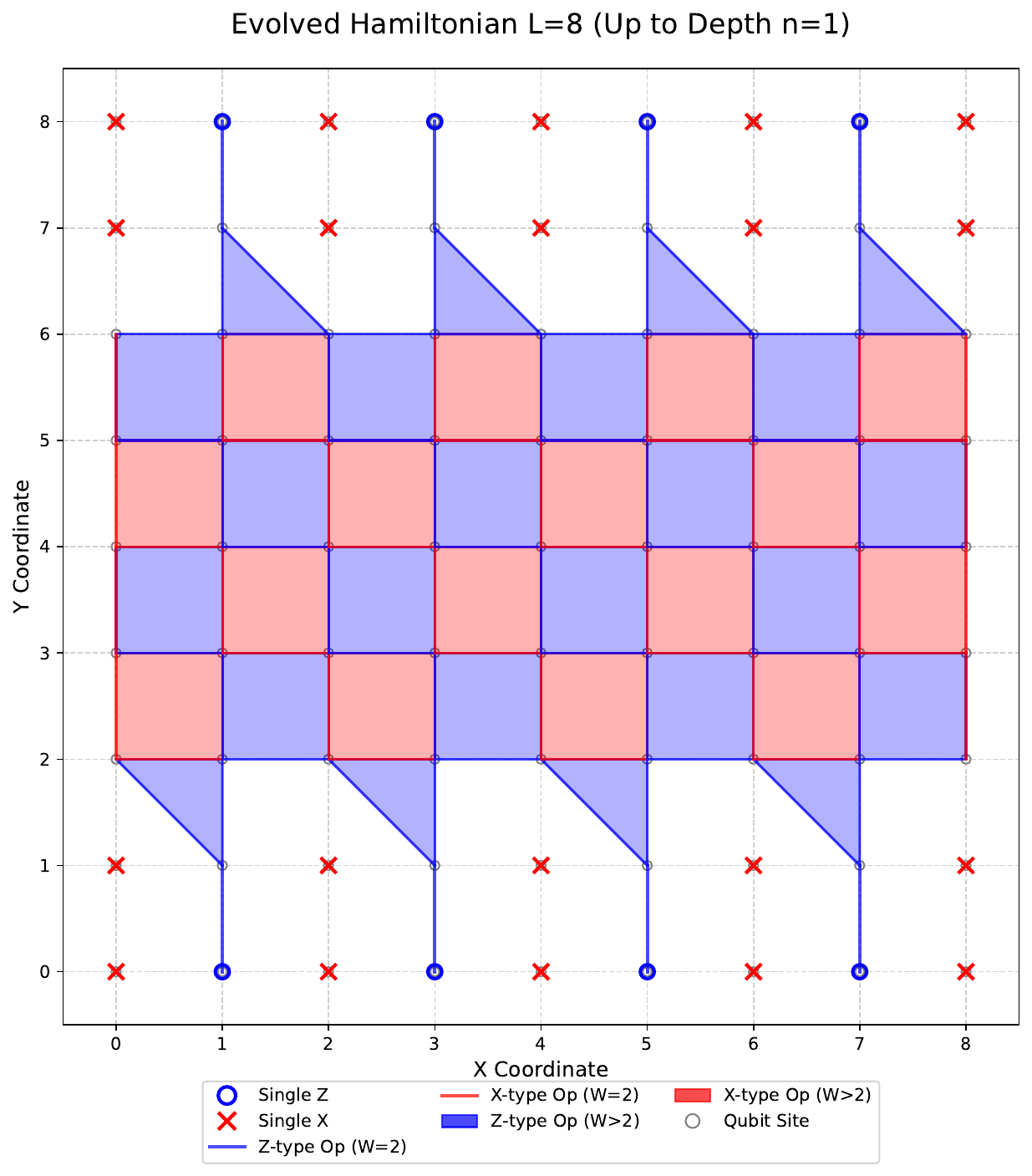}
    \caption{Resulting Hamiltonian after the second step of quantum circuit evolution}
    \label{fig:Ham2ndstepRSC}
\end{figure}

The next layer makes use of the fact that we obtain a copy of the first row of operators when we reflect by $x=L/2$, and hence reflecting the application of gates will result in the same form for the second to last row, with the $Z$ operators in the first row further reduced to only having support on 2 sites. The pattern continues, and all of the rows of $Z$ operators will be reduced to vertical 2-site operators at $x=1,3,..,L-1$. See the Hamiltonian obtained after the second step of the evolution in Fig. \ref{fig:Ham2ndstepRSC}.

\begin{figure} 
    \centering
    \includegraphics[width=8cm]{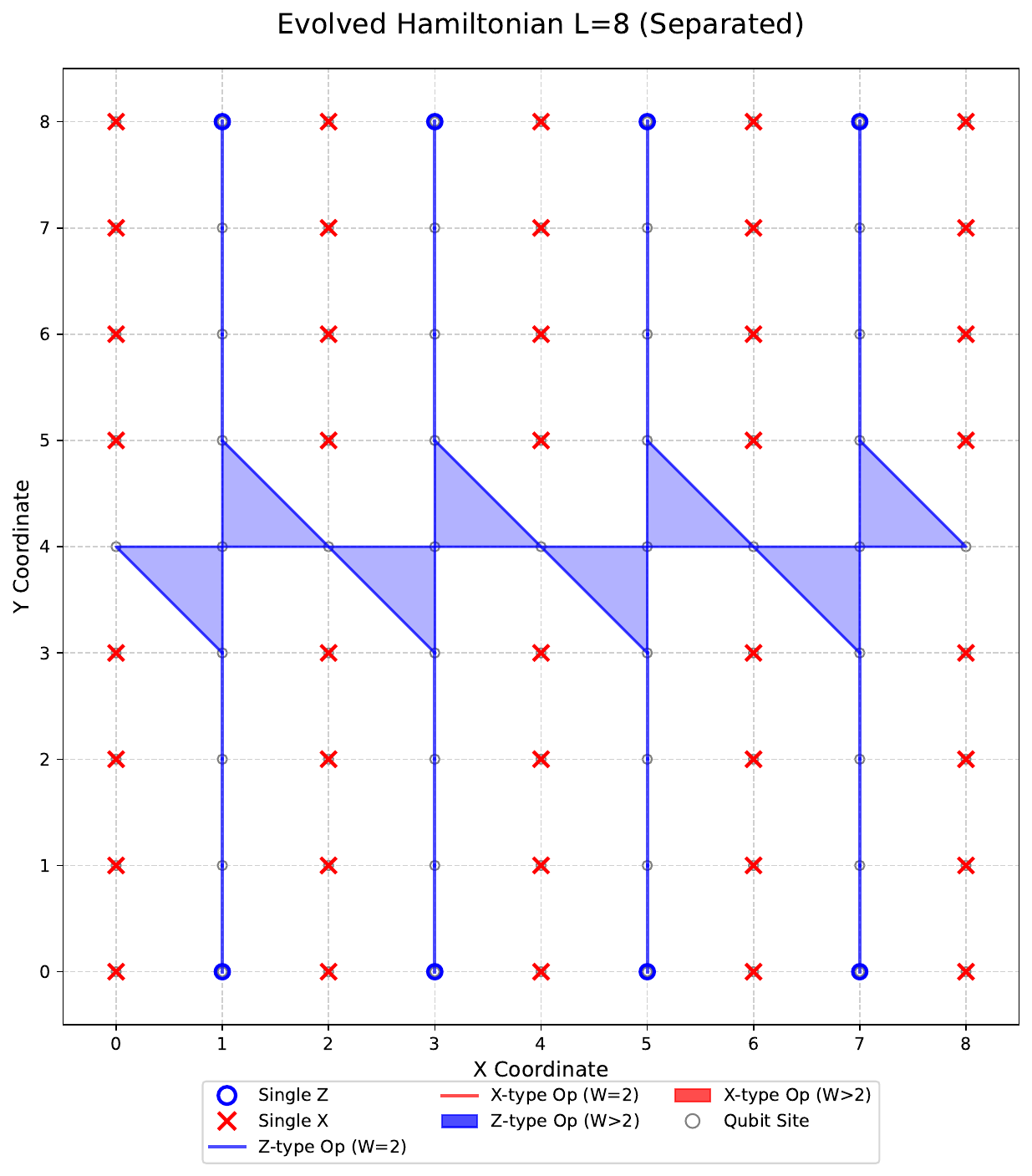}
    \caption{Resulting Hamiltonian after quantum circuit evolution.}
    \label{fig:HamfinalstepRSC}
\end{figure}

Finally the two rows of $Z$ operators in the middle are only reduced to 3-site operators, since they have only been reduced once. Then there is a $X$ local operator at sites $\{(2n, y): n\in\{0,1,..,L/2\}, y\in\{0,1,..,L\}\backslash \{L/2\}$, which is decoupled from the $Z$ subsystem. The resulting Hamiltonian is represented in Fig. \ref{fig:HamfinalstepRSC}.

We further note that the Z logical operator, originally vertically supported on qubits with $x=L/2$, is reduced to single site; this means that we can prepare the logical $\ket{0_L}$ state using a further circuit of depth one, by applying a Hadamard to all sites with $X$ qubits. However, we cannot prepare the logical $\ket{+_L}$ with a circuit of constant depth, since the logical $X$ operator is still spread over $L+1$ sites, with $y=L/2$, and the topological protection of this logical state still persists, as it is mapped to
\begin{equation*}
 \frac{1}{\sqrt{2}}\ket{+}^{\otimes L^2/2+L}\otimes \ket{0}^{\otimes L^2}\otimes (\ket{0}^{\otimes L+1}+\ket{1}^{\otimes L+1}) \, ,   
\end{equation*}
which makes constant depth preparation impossible due to the bound shown in \cite{Bravyi2006}. This is what we mean by encoding only one of the two logical operators: only the $Z$ eigenstates can be prepared with a circuit of constant depth after running the first half of this circuit. However note that there is a clear circuit of depth 1 to prepare $\ket{+_L}$ from $\ket{0_L}$: by applying a Hadamard onto all qubits. 

\begin{figure}[h!] 
    \centering
    \includegraphics[width=8cm]{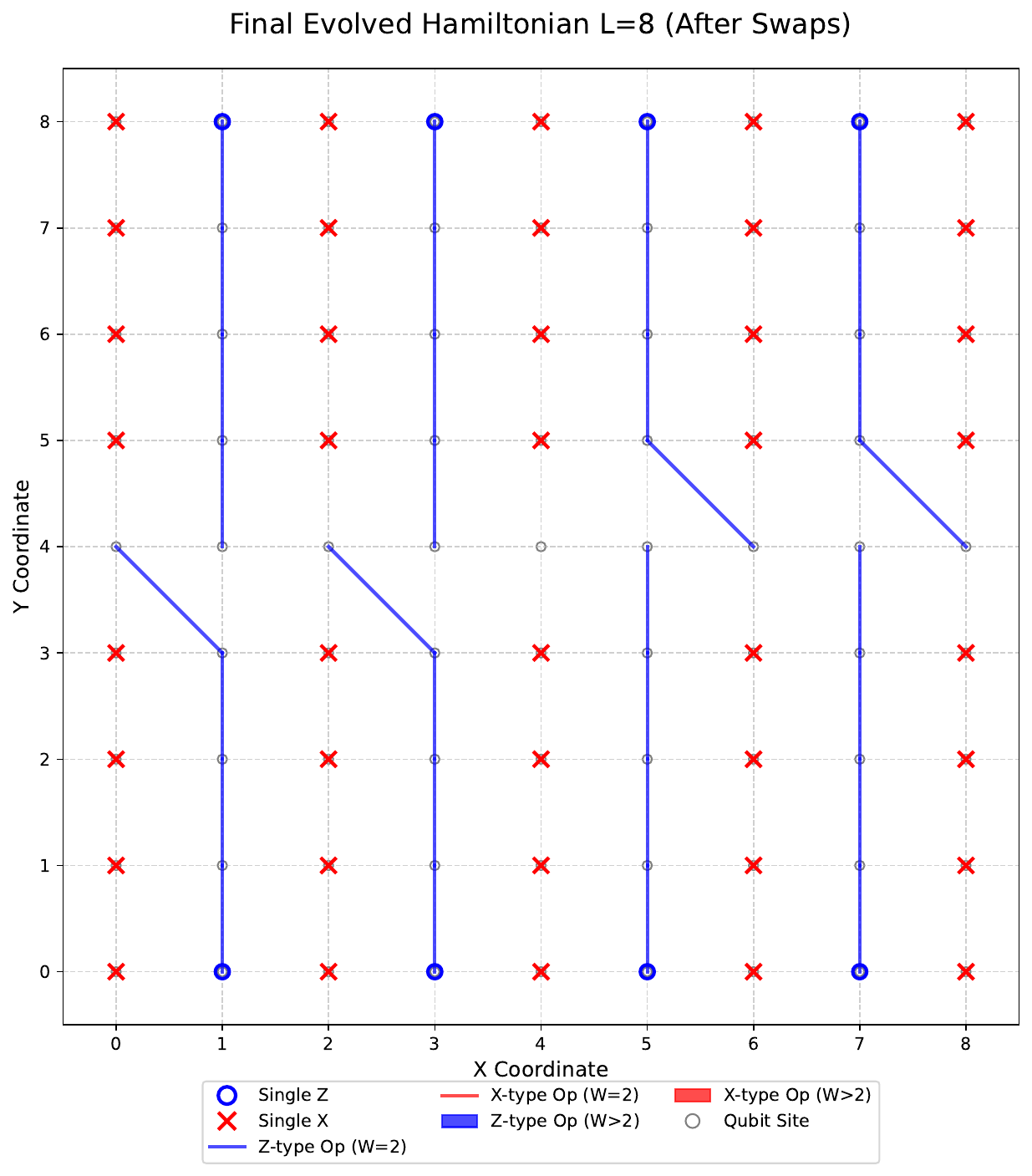}
    \caption{Resulting Hamiltonian after classical circuit evolution.}
    \label{fig:HamclassRSC}
\end{figure}

We can then further apply $XOR((j-1, L/2),(j, L/2))$ and $XOR((L-j+1, L/2),(L-j, L/2))$ at timestep $L/2+j$ for $j=1,..,L/2$, to result in a Hamiltonian that is of the form of $L$ Ising chains over $L+1$ qubits with a magnetic field in one end, plus $(L^2+2L)/2$ single site $X$ terms. This can be visualised in Fig. \ref{fig:HamclassRSC}.

It is not difficult to see why reducing a maximally independent set of operators of the same type (here without loss of generality, $X$ operators) to single site using $CX$ gates only, is sufficient to decouple the Hamiltonian into two disjoint subsystems: since the image of any $Z$ operator is still a $Z$ operator, and it must commute with all of the $X$ operators, it cannot be supported on any of the sites containing a lone $X$ operator. Since the remaining non-independent $X$ operators are also only supported on the sites with lone $X$, a $Z$ operator must commute with any $X$ operator trivially. This form is then simple to sample from classically, as discussed in Appendix \ref{appendix}. 

\begin{figure}[h!] 
    \centering
    \includegraphics[width=8cm]{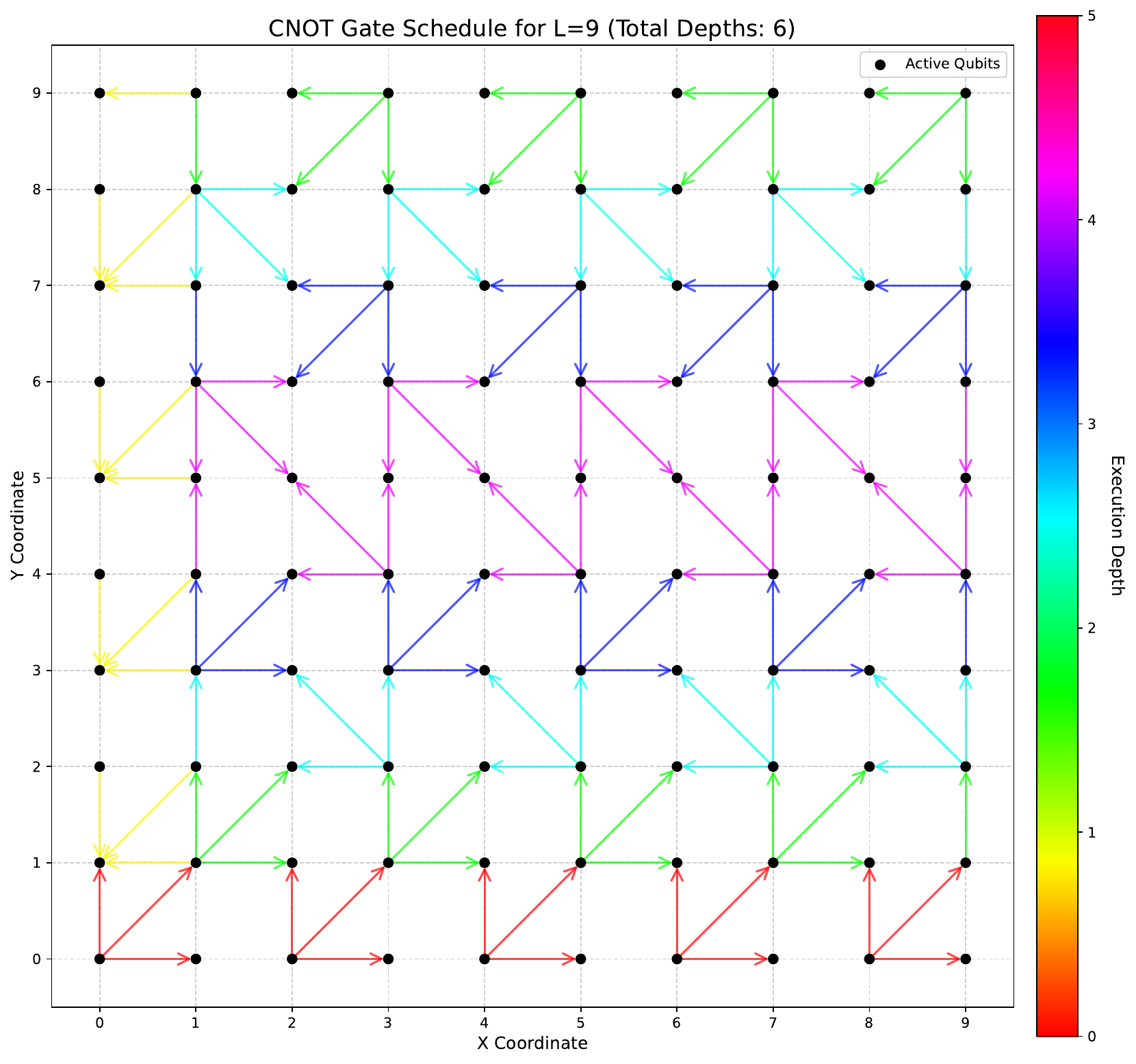}
    \caption{Gates used when $L=9$. Note the first two steps (red, yellow) are used to reduce to a smaller rotated surface code and is $O(1)$ depth. }
    \label{fig:RSCodd}
\end{figure}

\begin{figure}[h!] 
    \centering
    \includegraphics[width=8cm]{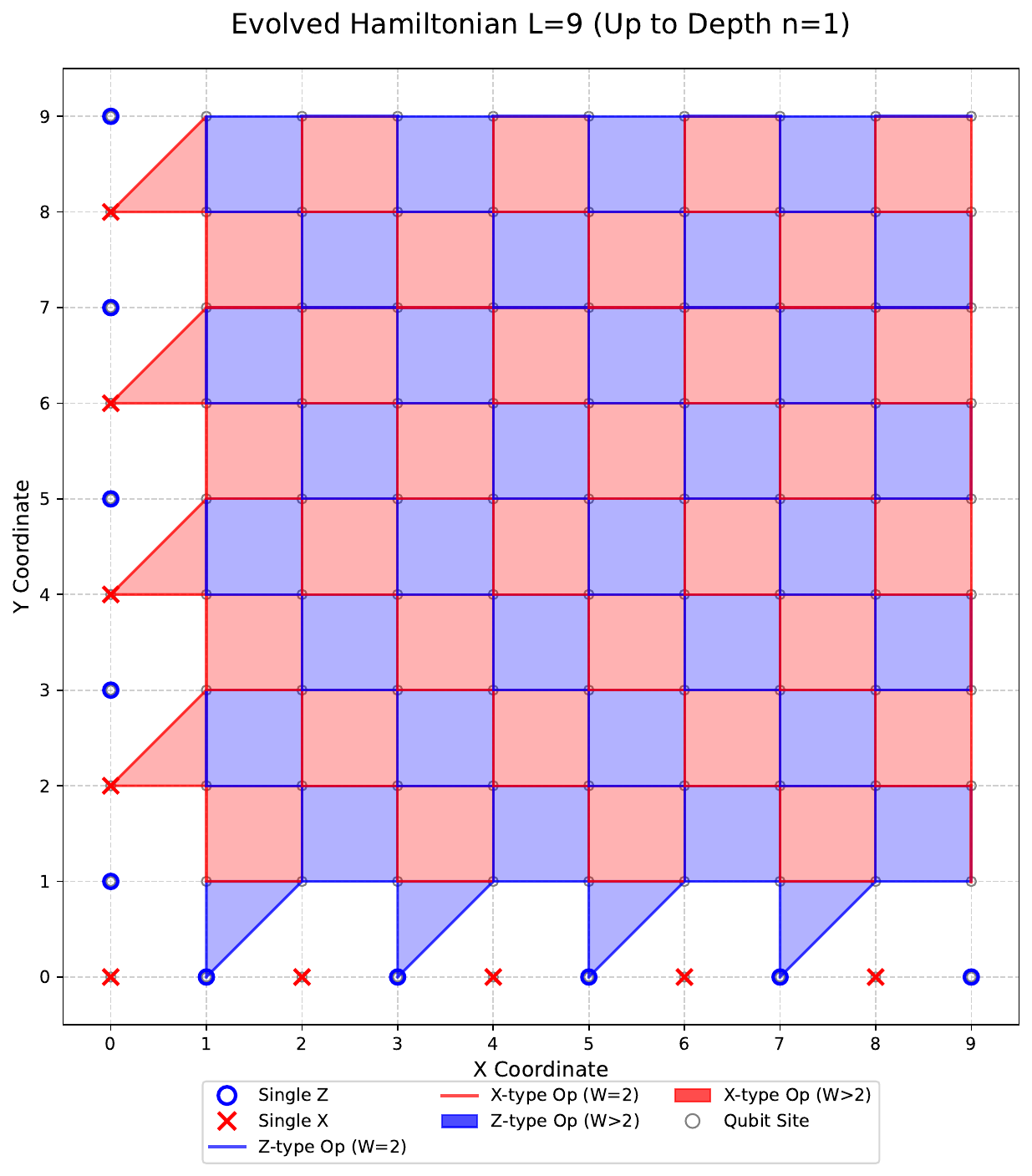}
    \caption{Resulting Hamiltonian after reducing to the even case.}
    \label{fig:RSCodd_redeven}
\end{figure}

\begin{figure}[h!] 
    \centering
    \includegraphics[width=8cm]{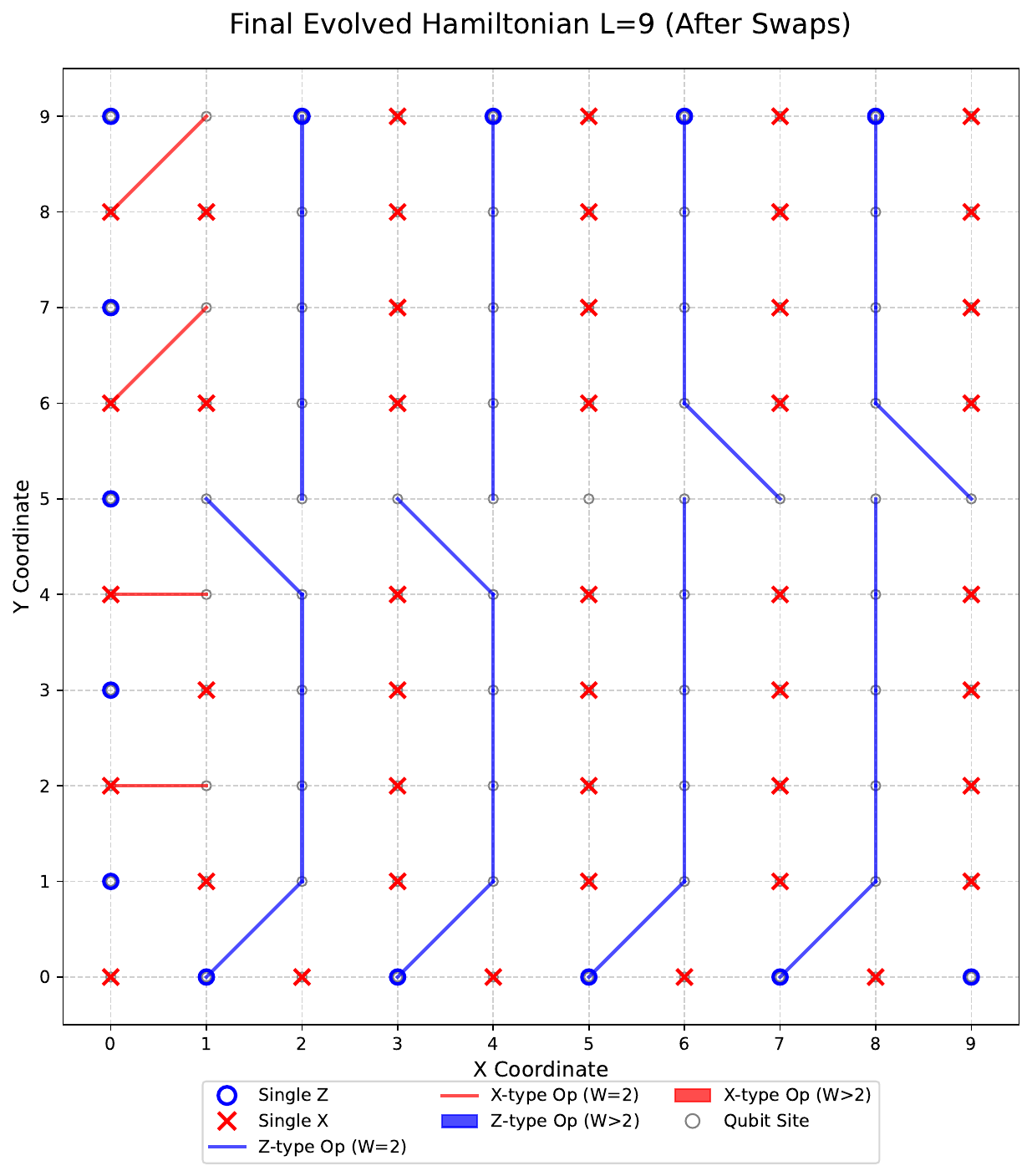}
    \caption{Resulting Hamiltonian after both quantum and classical circuit evolution: there are now $(L-1)/2$ Ising chains with a magnetic field at the end, over both $(L-1)/2$ and $(L+1)/2$ qubits. }
    \label{fig:RSCodd_final}
\end{figure}

Dealing with the case where $L$ is odd as well, we can use a circuit of depth 2 to reduce the rotated surface code Hamiltonian $H_{RSC}(L)$ to a sum of single site terms plus a Hamiltonian mostly trivial on the lines $x=0, y=0$ that resembles $H_{RSC}(L-1)$ on the $(L-1)\times (L-1)$ sub-lattice on the upper right hand corner, with only the boundary conditions along the left and bottom edges interacting with the qubits on $x=0, y=0$. Then applying the previous algorithm to $H_{RSC}(L-1)$ completes the proof. See an example of the initial layer of gates used in Fig. \ref{fig:RSCodd}, and of the resulting Hamiltonian after reducing to the even case in Fig. \ref{fig:RSCodd_redeven}.  The final Hamiltonian after both the quantum and classical evolution is represented in Fig. \ref{fig:RSCodd_final}.

We note that as a byproduct, this also encodes the logical state of the rotated surface code: the $Z$ and $X$ logical operators that are originally vertical and horizontal respectively and go through the centre qubit, are reduced to a single site term at the centre. This is equivalent to encoding a logical state after a further depth 1 preparation of the ground state of the system, which is independent of the information stored in the centre qubit, as argued in \cite{Shum2025}; and the depth matches that of \cite{Claes2025}.

However, our idea here is special in the sense that when evolving only the quantum part of the circuit, i.e. up to timestep $L/2$, only the vertical $Z$ logical operator is reduced to single site at the middle; the horizontal $X$ logical operator is stabilized up to this point. Indeed, we can consider the further classical evolution as reducing the $X$ logical operator to single site as well. If one then agrees with reducing both logical operators to the same site and also decoupling the system as being 1 step away from encoding the logical state, we have essentially encoded only one of the 2 logical operators using quantum gates, and the other is encoded using classical gates.

\section{Hybrid Gibbs sampling the toric code}

\begin{figure}[h!] 
    \centering
    \includegraphics[width=8cm]{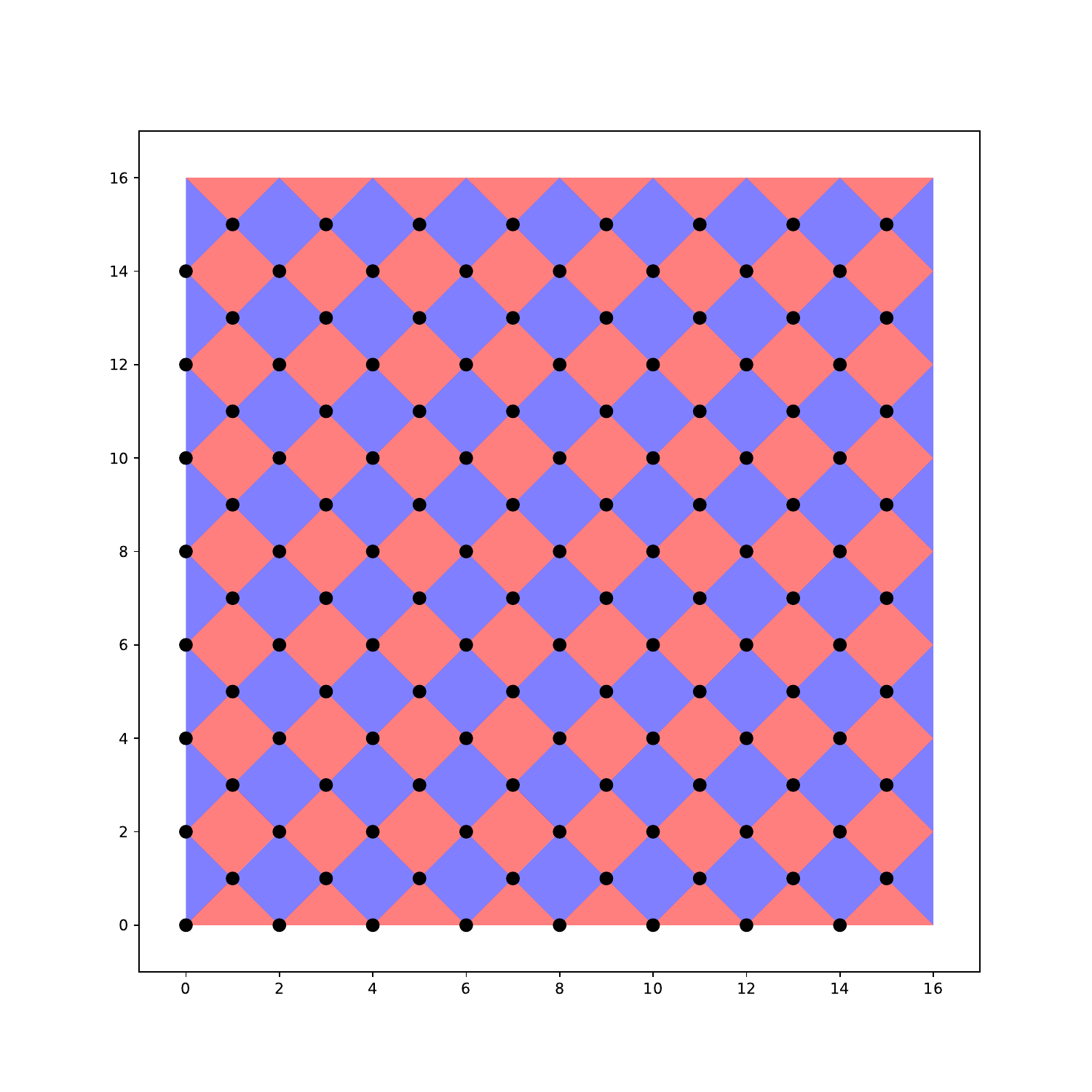}
    \caption{Toric code plot. Note that the triangle terms on the top/bottom and left/right boundaries combine to form one plaquette term under the toric topology. As before, $X/Z$ terms are represented with filled red/blue squares. }
    \label{fig:toric_code}
\end{figure}

A similar idea then applies for the 2D toric code, where we take $L$ depth to reduce all $Z$ operators to single site. Note that the construction is similar to that in \cite{Chen2024}; however, we also show that it is possible to prepare the logical $\ket{00_L}$ in $L+1$ depth using local gates, improving on the previous bound of $2L+2$. We choose an embedding of the toric code Hamiltonian with qubits placed on the coordinates $(x,y)\in \mathbb{Z}^2:x+y\equiv 0 \, (\hspace{-0.2cm}\mod 2), 0\leq x,y\leq 2L-1$ where arithmetic is done mod $2L$, with the $X$ plaquette terms centred at sites $(2i+1, 2j)$ and the $Z$ plaquette terms centred at sites $(2i, 2j)$. A representation of this Hamiltonian can be seen in Fig. \ref{fig:toric_code}. 

\begin{figure}[h!] 
    \centering
    \includegraphics[width=8cm]{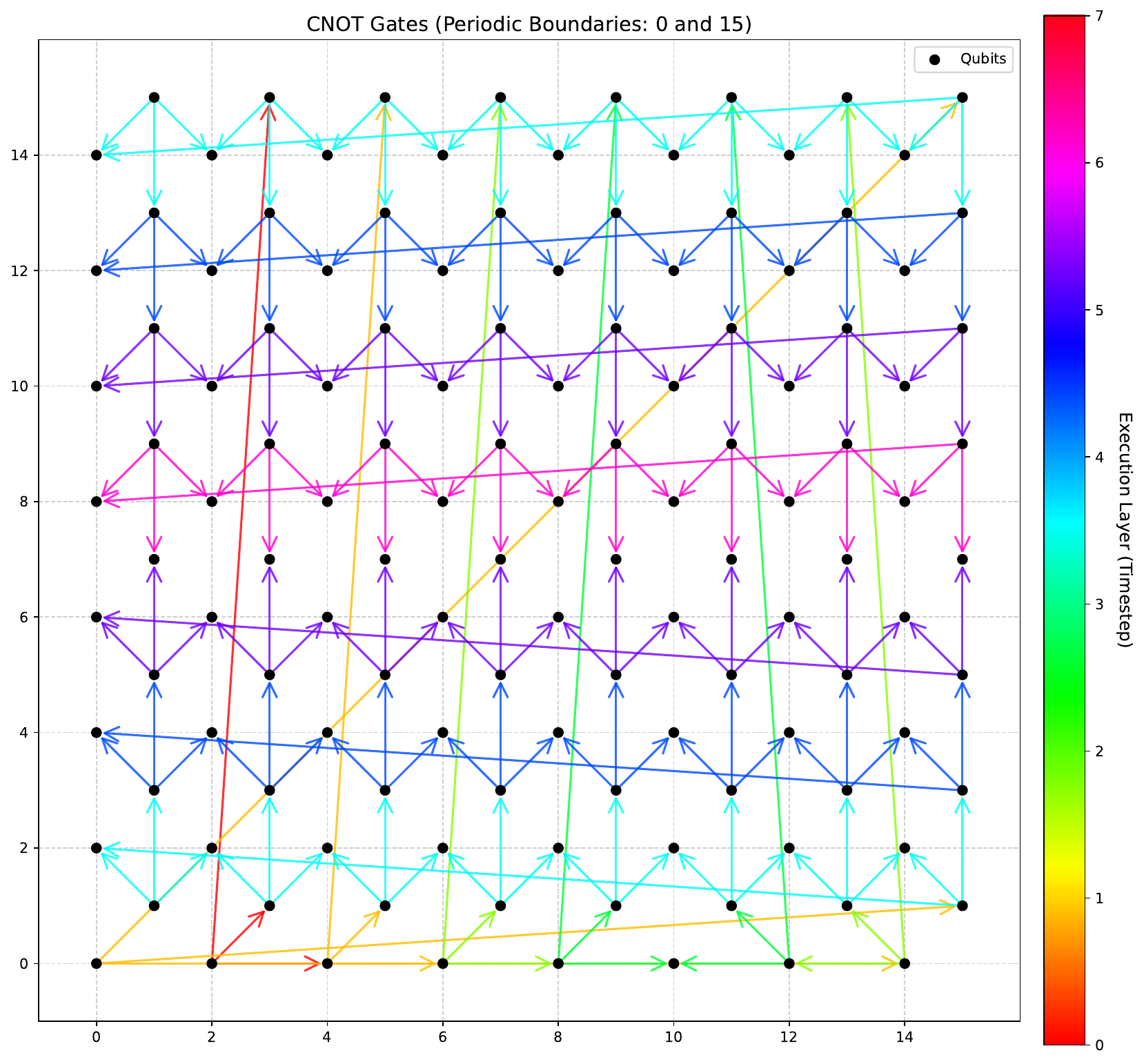}
    \caption{Gate set required to separate the Hamiltonian into 2 disjoint subsystems. Note the gates are still local since we are under the toric topology. }
    \label{fig:toric1ststep}
\end{figure}

\begin{figure}[h!]
    \centering
    \includegraphics[width=8cm]{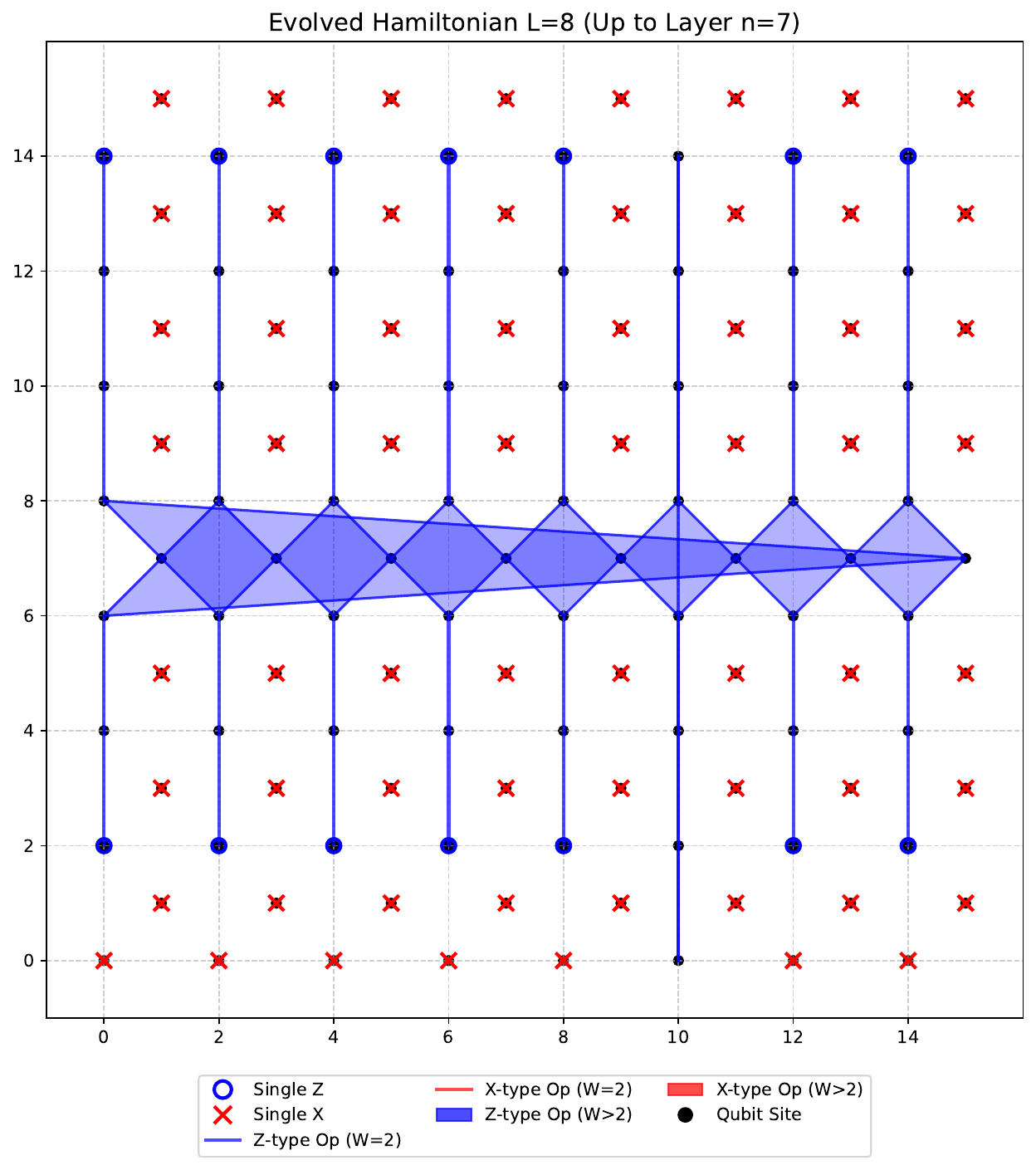}
    \caption{The toric code Hamiltonian after evolving through the separating gate set. We choose to plot all $Z$ operators, and $L^2-1$ of the $X$ operators, which have all been reduced to single site; the final $X$ operator is a product of the single site $X$ plotted. }
    \label{fig:toricHam1ststep}
\end{figure}

The algorithm reduces a horizontal logical $Z$ operator to single site in $\lceil L/2\rceil$ depth, then reduces a vertical logical $Z$ operator to single site in $\lfloor L/2 \rfloor$ depth; it simultaneously reduces $L^2-1$ $X$ operators to single site, with the final non-independent operator being a product of all these $L^2-1$ operators. This is represented in Fig. \ref{fig:toric1ststep}, and the resulting Hamiltonian can be seen in Fig. \ref{fig:toricHam1ststep}.

Then we can prepare the logical state $\ket{00_L}$ by noting that it has been mapped to the product state: the $X$ logical operators originally with support on $x = \begin{cases}
L+2, & L \text{ even}\\
L+1, & L \text{ odd}
\end{cases}$, say $X_1$, and $y = \begin{cases}
L-1, & L \text{ even}\\
L, & L \text{ odd}
\end{cases}$, say $X_2$, are stabilized after the separating evolution; while the $X$ logical operators evolve to any single site term on these vertical and horizontal lines (up to stabilizer gauge). Hence the $\ket{00_L}$ state here, defined as the simultaneous ground state of all stabilizers and $Z_1, Z_2$,  is mapped to the product state of setting all qubits with a $X$ term on it to $\ket+$, and setting all qubits with a $Z$ term on it to $\ket0$, which can be prepared with a layer of $H$ gates. Hence we require depth $L+1$ to prepare $\ket{00_L}$ from $\ket{0}^{\otimes 2L^2}$ using local gates, improving on the previous result by \cite{Chen2024} which required depth $2L+2$. 

As argued in the section on the rotated surface code, we cannot prepare the logical $\ket{++_L}$ directly. However, we can always prepare a $\ket{00_L}$ and apply a layer of Hadamards, achieving a circuit depth of $L+2$; on the contrary, it is impossible to prepare the logical Bell states, defined as the simultaneous eigenstates of $X_1X_2$ and $Z_1Z_2$ with a further constant depth local circuit. Indeed, to prepare these states with the ideas presented above, we will have to reduce these minimum weight $d=2L$ operators to single site, which cannot be done using local gates and a depth of $\sim L$. 

Since we have separated the two systems, the evolution is now completely classical (after a layer of $H$ gates on all qubits holding an $X$ implied), and we drop the locality requirements; 1 more layer of $XOR$ operations will give us two decoupled Hamiltonians, which are both unitarily equivalent to an Ising chain with a magnetic field on both ends, and hence can be Gibbs sampled from in $\mathcal O(L^2)$ classical operations. 

\begin{figure} 
    \centering
    \includegraphics[width=8cm]{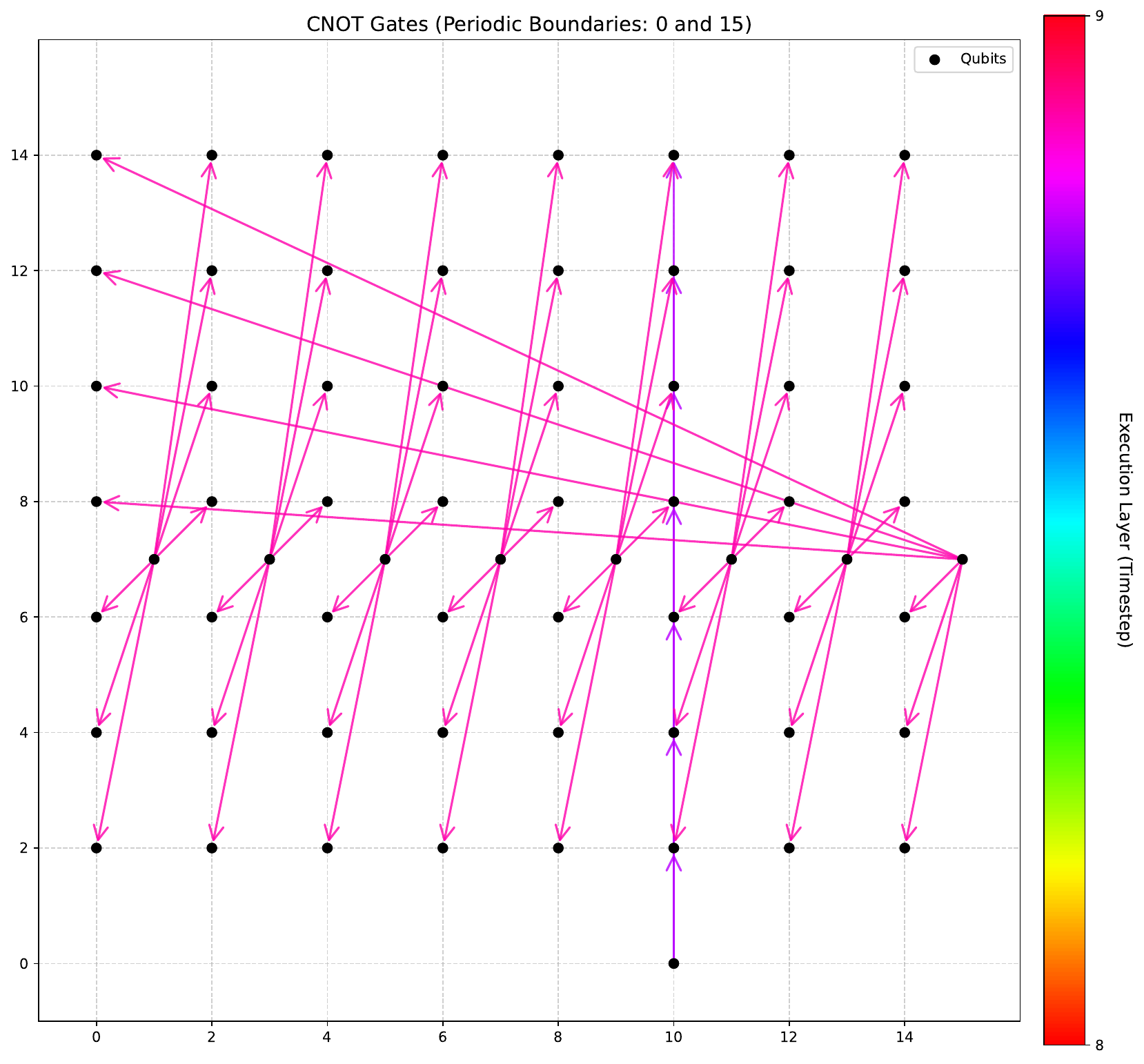}
    \caption{The final layer of commuting $XOR$ to be applied onto the $Z$ subsystem. Note the colour difference is only due to the different nature of the two gatesets: the purple and pink gatesets reduce the two logical $X$ operators to single site. }
\end{figure}

\begin{figure} 
    \centering
    \includegraphics[width=8cm]{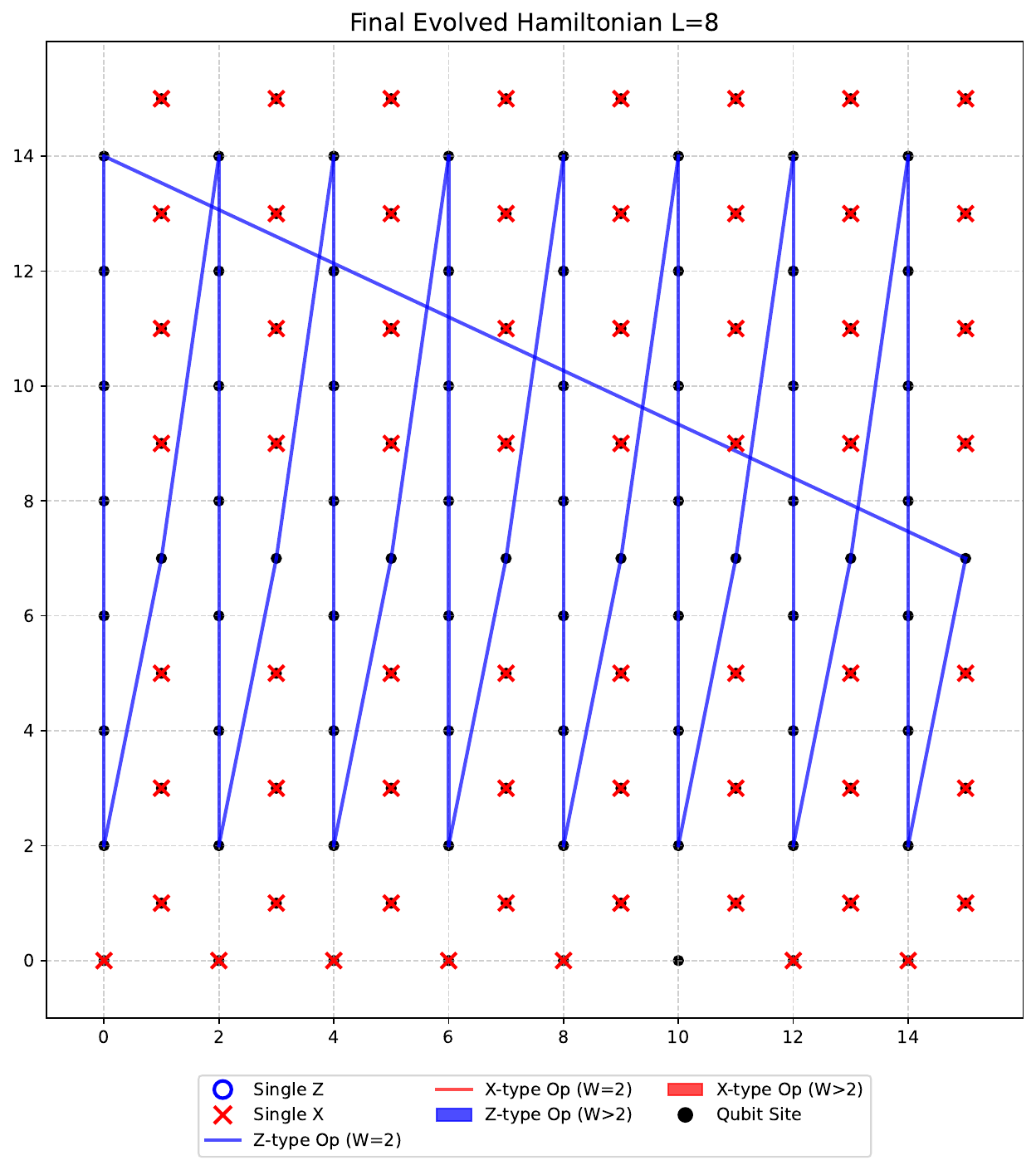}
    \caption{Final form of the Hamiltonian. The $X$ subsystem is of the form $H_X=-\sum_{i=1}^{L^2-1} X_i-\Pi_{i=1}^{L^2-1} X_i$, defined over $L^2$ qubits, called the parity check Hamiltonian, where the non-independent product term is not plotted, and the $Z$ subsystem is an Ising loop over $L^2$ qubits $H_Z=-Z_{L^2}Z_1-\sum_{i=1}^{L^2-1}Z_iZ_{i+1}$. }
\end{figure}

We then observe that in principle, the sampling step of the rotated surface code can be performed with constant depth using qubit measurements, since all stabilizers are independent and hence the Gibbs state tensorises, while we need depth scaling with the size of the system for the toric code, since the probability weights cannot be split into the product of independent random variables, and hence require other nondirect sampling methods. 

\section{Quantum Gibbs sampling the Ising loop}

To consider how optimal our algorithm is, we note that a sampling algorithm should be run many times, and if we do so on the same qubits, no matter what state and what further manipulation we perform, after terminating by reading a measurement of $\mathcal O(N)$ independent operators, $N$ being the number of qubits, we need to reset it to $\ket{0}^{\otimes N}$ to run it again, which takes $\mathcal O(N)$ time to do so. Hence it remains unnecessary to come up with a sampling algorithm that runs quicker than $\mathcal O(N)$. 

Nevertheless, we provide a non-local quantum algorithm here to sample from the parity check Hamiltonian 
\begin{equation*}
    H_{n+1}=-\sum_{i=1}^{n+1} Z_i-\Pi_{i=1}^{n+1} Z_i \, ,
\end{equation*}
that is to prepare 
\begin{equation*}
  \rho_{n+1}\propto \exp \left(\beta\sum_{i=1}^{n+1} Z_i+\beta\prod_{i=1}^{n+1} Z_i\right) \, , 
\end{equation*}
defined over $n+1$ qubits; we decompose the circuit into $Y$ rotations, which are non-Clifford and will take two different rotation angles, and $H, X, CX$ gates. The whole circuit can be executed in $\mathcal O(1)$ depth, combined with measurements on at most $n+2$ qubits, where $n$ of which can be performed simultaneously. We stress that it is possible to simultaneously measure $N$ qubits in $\mathcal O(1)$ time, say in the computational basis, but we cannot read the result in $\mathcal O(1)$ time; that should take $\mathcal O(N)$ in general. However, since we are preparing a mixed state, it is crucial to differentiate between just measuring the result and also reading it, e.g. as in \cite{schmidhuber2025hamiltoniandecodedquantuminterferometry}. Note that this Hamiltonian defined over $n+2$ qubits is unitarily equivalent to the 1D Ising loop, and hence two disjoint copies of this Hamiltonian, defined over $2(n+2)=2L^2$ qubits, is unitarily equivalent to the toric code. This is where the idea will be useful, as to be covered in the next section: we impose locality to obtain an algorithm that Gibbs samples the toric code in $\mathcal O(L)$ depth of one and two qubit gates, combined with $\sim 2L^2$ measurements, of which $2L^2-2$ can be performed simultaneously. 

We first implement the classical Gibbs sampling idea over qubits: it is clear that if we start with a mixed state, we can perform Gibbs sampling without reading the state using $CX$ gates. Explicitly, we provide an algorithm that prepares $\rho_{n+1}$ given $\rho_n\otimes \ket{+1}\bra{+1}$. Note that under the distribution $\rho_{n+1}$, conditioned on the first $n$ qubits being in computational state $\ket{x}$, where we label the state using $Z$ eigenvalues instead of the logical values, i.e. $\pm1$ instead of $\{0,1\}$, the relative probabilities of $\ket{x}\ket{+1}$ and $\ket{x}\ket{-1}$ only depend on the parity of $x$: 
\begin{equation*}
  \dfrac{\mathbb{P}(\ket{x}\ket{+1}|x)}{\mathbb{P}(\ket{x}\ket{-1}|x)}=\exp(2\beta(1+\pi(x)))   \, ,
\end{equation*}
 where $\pi(x)$ is the parity of $x$ taking values $\pm1$. Hence as long as we can compute the parity of a $n$ qubit unknown computational basis state efficiently, e.g. using $U=\prod_{i=1}^nCX(i, n+2)$ on an ancilla qubit number $n+2$ set to $\ket{+1}$ originally, and apply a controlled rotation controlled by this parity value to the qubit $n+1$ such that it measures to the correct statistics and then measure it, discarding the $n+2$th qubit, we have prepared $\rho_{n+1}$ from $\rho_n\otimes \ket{+1}\bra{+1}$. The controlled rotation required, say $R$, rotates $\ket{+1}\ket{+1}$ to 
 \begin{equation*}
   \dfrac{1}{(e^\beta+e^{-\beta})^{1/2}}(e^{\beta/2}\ket{+1}\ket{+1}+e^{-\beta/2}\ket{+1}\ket{-1})  
 \end{equation*}
and $\ket{-1}\ket{+1}$ to 
\begin{equation*}
    \dfrac{1}{\sqrt{2}}(\ket{-1}\ket{+1}+\ket{-1}\ket{-1}) \, .
\end{equation*}
The whole circuit can be composed in constant depth, so if we implement the classical algorithm directly using qubits we will need $\mathcal O(n)$ depth for an Ising loop of length $n+2$. 

We now consider the parity check term $-\Pi_{i=1}^n Z_i$ as an energy splitting term, perturbing the energy levels but not the eigenstates of the magnetic field term $-\sum_{i=1}^n Z_i$. It is then possible to sample from the magnetic field term in $\mathcal O(1)$ depth and a measurement of all $n+1$ qubits; starting in $\ket{+1}^{\otimes n+1}$, rotate qubits $1,2,..,n+1$  by $\exp(-iY\theta)$, with $\theta=\tan^{-1}(e^{-\beta})$ which maps 
\begin{align*}
    \ket{+1}& \rightarrow\cos\theta\ket{+1}+\sin\theta\ket{-1}:=\ket{\psi} \, , \\
    \ket{-1}&\rightarrow-\sin\theta\ket{+1}+\cos\theta\ket{-1} \, .
\end{align*}
We then measure all first $n+1$ qubits simultaneously, yielding the Gibbs state of $-\sum_{i=1}^{n+1} Z_i$ at inverse temperature $\beta$. 

We then deal with the parity check term: here we start requiring that $n$ is even. Note that this can be done without loss of generality: if we want to prepare the Gibbs state for odd $n$, we can simply prepare the Gibbs state $\rho_{n-1}\otimes\ket{+1}\bra{+1}$ and extend it to $n$ qubits with the classical method described above. Preparing $\ket{\psi}^{\otimes n+1} \ket{+1}$ as above, we want a unitary $U$ that computes the eigenvalue of the computational basis state defined over the first $n+1$ qubits. That is, defining $\ket{\Psi}=\ket{\psi}^{\otimes n+1}, $let$ \ket{\Psi}=\sqrt{\pi_0}\ket{\Psi_0}+\sqrt{\pi_1} \ket{\Psi_1}$, with $\pi_0=\dfrac{1+\tanh^{n+1}\beta}{2}, \pi_1=\dfrac{1-\tanh^{n+1}\beta}{2}$ the respective probabilities of obtaining an even/odd state after performing a Gibbs sample of the magnetic field Hamiltonian, $\ket{\Psi_0}, \ket{\Psi_1}$ the normalised projections of $\ket{\Psi}$ onto the $+1, -1$ eigenspaces of $-\Pi_{i=1}^{n+1} Z_i$ respectively, we want 
\begin{align*}
    U_{n+1}: (\mathbb{C}^2)^{\otimes n+2} \rightarrow (\mathbb{C}^2)^{\otimes n+2}  \, ,\\
    U\ket{\Psi_0}\ket{+1}=\ket{\Psi_0}\ket{+1} \, , \\
    U\ket{\Psi_1}\ket{+1}=\ket{\Psi_1}\ket{-1} \, .
\end{align*}
 Such a unitary exists and can be decomposed into 2 qubit gates, e.g. $U=\prod _{i=1}^{n+1} CX(i, n+2)$; however it is non local, and we will come back to this in the next section. 

To account for the Boltzmann weight due to this term, we apply $U_{n+1}$ to the state $\ket{\Psi}\ket{+1}$ to obtain $\sqrt{\pi_0}\ket{\Psi_0}\ket{+1}+\sqrt{\pi_1} \ket{\Psi_1}\ket{-1}$. We then measure the $n+2^{th}$ qubit, and read it. 

If we get $+1$, note that any further measurement of the first $n+1$ qubits gives an even parity state, and indeed measuring only the first $n$ qubits gives our desired Gibbs state: denoting a computational basis state by its $Z$ eigenvalues, the probability of obtaining $\ket{z_1,z_2,..,z_{n+1}}$ after further measurement on the first $n+1$ qubits is 
\begin{equation*}
   \mathbb{P}(z_1, ..,z_{n+1})=\begin{cases}
0, & x \text{ odd}\\
\dfrac{ \exp(+\beta\sum_{i=1}^{n} z_i+\beta\Pi_{i=1}^{n} z_i)}{2^n (\sinh^{n+1}\beta+\cosh^{n+1}\beta)}, & x \text{ even} \, ;
\end{cases} 
\end{equation*}
 since given the first $n$ qubits in some computational state, there is exactly one state on $n+1$ qubits that have even parity and match on the first $n$ qubits, the state obtained after measuring the first $n$ qubits is the Gibbs state $\rho_{n}$, while collapsing the $n+1$th qubit to a computational basis state; applying a Hadamard and measuring it gives the maximally mixed state on that qubit, and we have prepared $\rho_n\otimes \mathbb{I}/2$ on the first $n$ qubits; classical Gibbs sampling takes us to $\rho_{n+1}$. 

Else we get $-1$, and the partials on the first $n$ qubits instead measure to the density matrix
\begin{equation*}
    \rho\propto \exp \left(\beta\sum_{i=1}^{n} Z_i-\beta\prod_{i=1}^{n} Z_i\right) \, ;
\end{equation*}
 measuring the first $n$ qubits as before, we apply an $X$ gate onto all first $n$ sites; using the arguments presented above, the density matrix evolves as $\rho$ mentioned above, where the sign of the product term does not flip since $n$ is even, and it is the Gibbs state of the Hamiltonian $H_n$ at negative temperature $-\beta$. Again the $n+1$th qubit collapses to an unknown computational basis state, so we can apply a Hadamard and measure it to obtain
 \begin{equation*}
    \rho\propto \exp \left(-\beta\sum_{i=1}^{n} Z_i-\beta\prod_{i=1}^{n} Z_i\right) \otimes\mathbb{I}\, ;
\end{equation*}
 over the first $n+1$ qubits. Note that the classical procedure works as well for negative temperatures, and hence we can convert the state to $\rho\propto \exp \left(-\beta\sum_{i=1}^{n+1} Z_i-\beta\prod_{i=1}^{n+1} Z_i\right)$. Further applying an $X$ on all $n+1$ sites gives us $\rho=\rho_{n+1}$, since $n+1$ is odd. 

Finally note that if we apply a non-local circuit of depth $\lceil{\log_2n}\rceil$ we can convert the single site term to the 1D Ising chain with magnetic fields on both ends as follows: use a shorthand of $a_1, a_2, \ldots ,a_k$ over $\sum_{i=1}^ka_i=n$ qubits to represent a Hamiltonian defined over these qubits, as a sum of Ising chains starting from sites $1, 1+a_1, 1+a_1+a_2,..,1+\sum_{i=1}^{k-1}a_i$ of lengths $a_1, a_2, ..,a_k$ respectively, with a magnetic field at the start of the chain, plus a product term as the product of all previous terms. For example (3,1,2) means 
\begin{equation*}
    -Z_1-Z_1Z_2-Z_2Z_3-Z_4-Z_5-Z_5Z_6-Z_3Z_4Z_6 \, ,
\end{equation*}
 defined over $3+1+2=6$ qubits. Then our initial Hamiltonian is $(1,1,..,1)$ with $n$ entries, and we use the gateset below to join two Ising chains with a magnetic field in front of arbitrary length: 
 \begin{equation*}
     H=-Z_1-\sum_{i=1}^kZ_iZ_{i+1}-Z_{k+2}-\sum_{i=k+2}^mZ_iZ_{i+1}
 \end{equation*}
 is mapped to $H'=-Z_1-\sum_{i=1}^{m-1}Z_iZ_{i+1}$ by $\prod_{i=k+2}^l CX(k+1, i)$ which is a depth 1 product of 2 qubit gates. Hence we can convert $(1,1,1,..,1)$ to $(n)$ in $\lceil \log_2n\rceil$ depth by applying the above idea recursively. The final term falls into place as a magnetic field at the end of the chain since unitary conjugation preserves the fact that the final term is the product of all previous terms. Finally, resetting the $n+2^{th}$ qubit to $\ket{+1}$ and applying $\prod_{i=1}^{n+1}CX(n+2, i)$ gives the Gibbs state of the Ising loop over $n+2$ qubits. Hence we can Gibbs sample the $n+2$ qubit Ising chain in an $\mathcal O(\log n)$ depth circuit of one and two qubit gates.

 \begin{figure}[h!] 
    \centering
    \includegraphics[width=8cm]{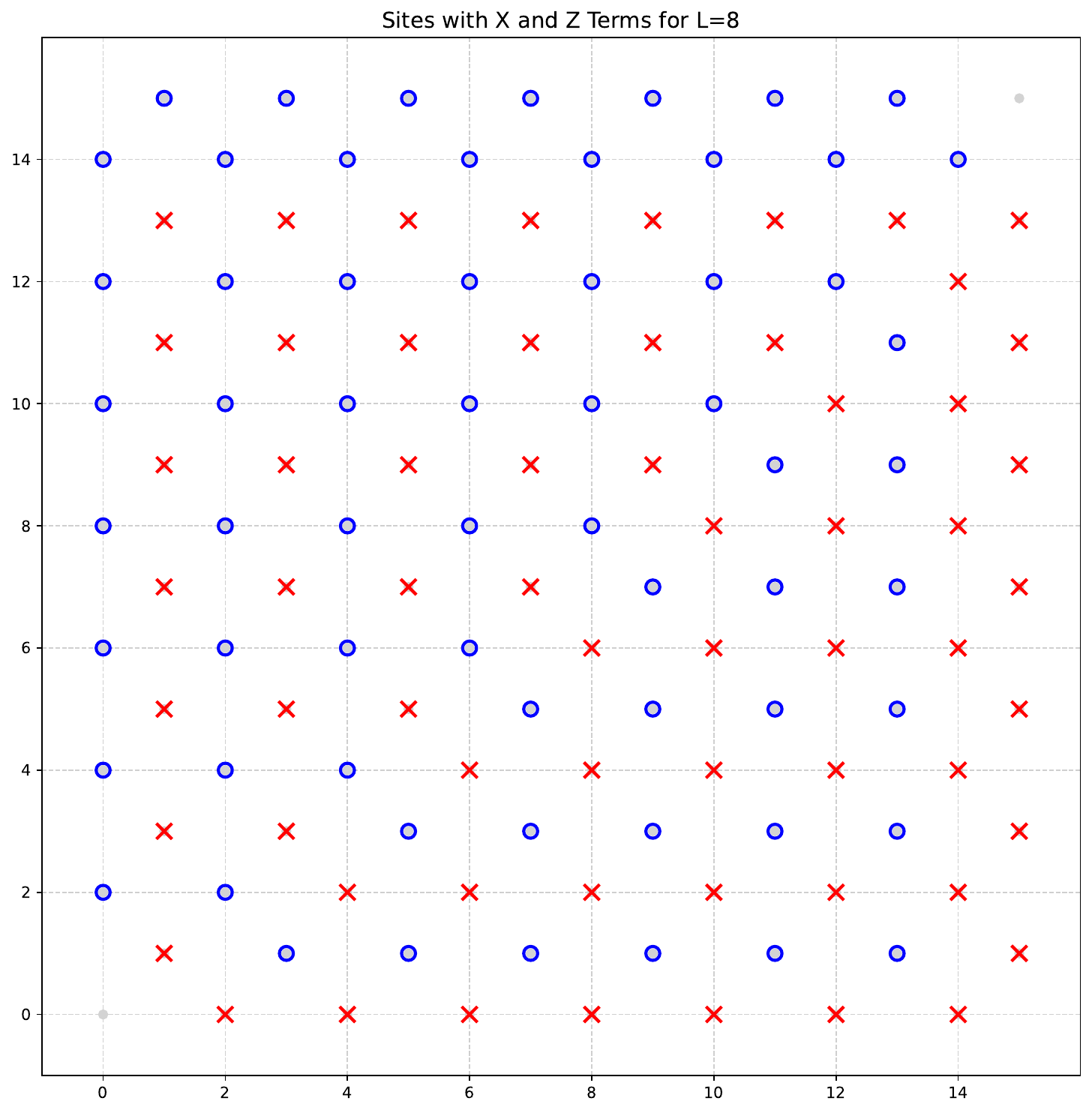}
    \caption{The parity check Hamiltonians after evolving the toric code Hamiltonian using gateset $W$ (before applying the final layer of Hadamards). The $2L-3$ depth psuedo parity check can then be executed on the two subsystems separately, where for the $Z$ and diagonalised $X$ subsystems, we start from the bottom/right, and the $CX$ arrow points up/left to the nearest qubit of same type; after all of the parity information is collected at the top/left line of qubits, we point right/down to the ancillae qubits at the top right/lower left respectively. }
    \label{fig:HamiltonianGS}
\end{figure}

\section{Quantum Gibbs sampling the toric code}

\begin{figure}[h!] 
    \centering
    \includegraphics[width=8cm]{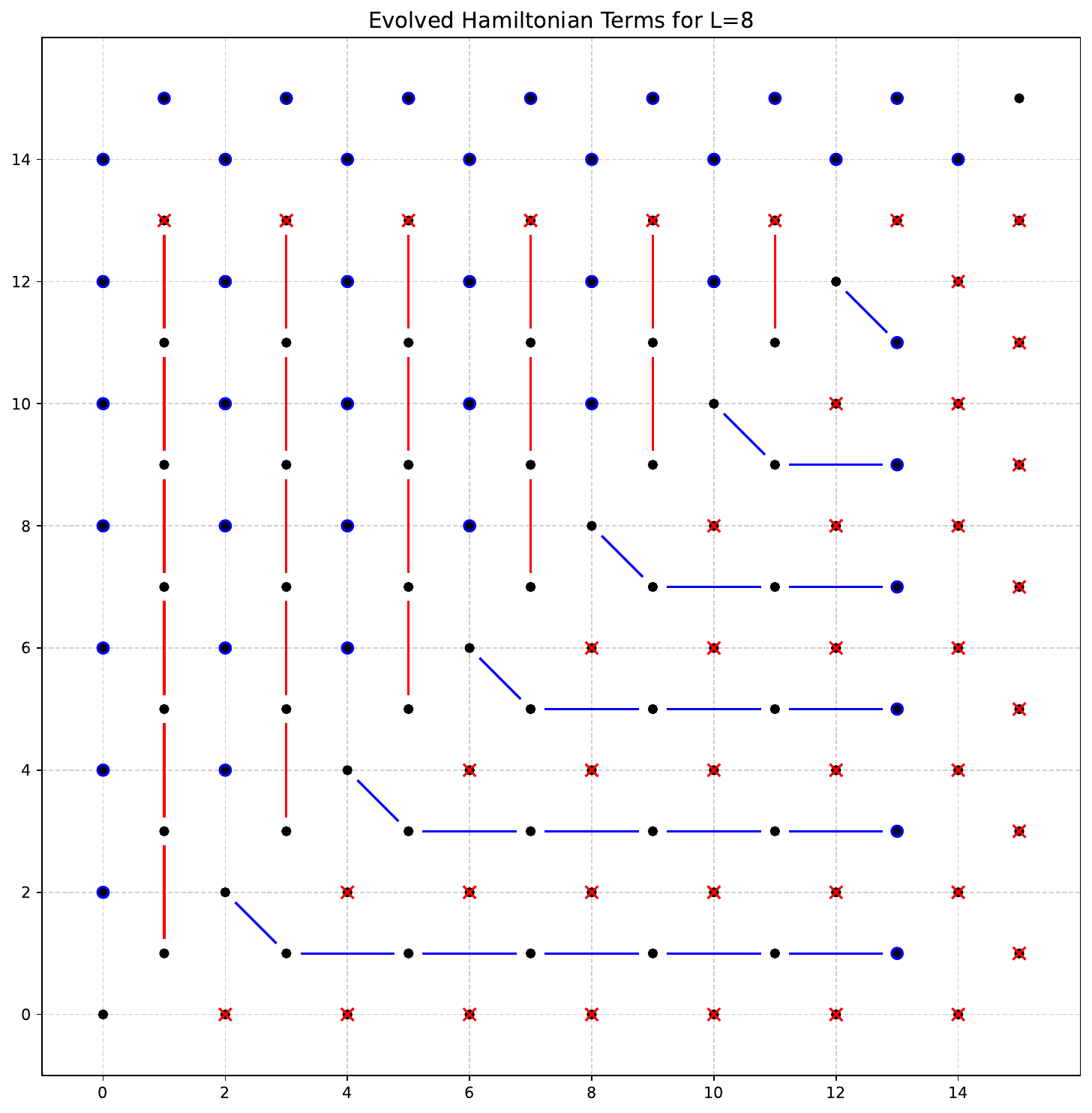}
    \caption{The final form in the preparation of the ground state in \cite{Shum2025}. We have only plotted the independent terms; the last term of both $Z$(blue) and $X$(red) type is the product of all the $Z/X$ terms plotted. A further depth $L-2$ layer of $CX$ gates along the Ising chain terms (for $X/Z$ terms, start from bottom/left pointing upwards/right for each chain) will reduce the Hamiltonian to desired form, and a further depth 1 layer of $H$ gates on the $X$ qubits diagonalises it. }
    \label{fig:circuitGS}
\end{figure}

To use the result above, we need to put the toric code in decoupled parity check form. Indeed, it was shown in \cite{Shum2025} that we can prepare the ground state in $2L+1$ depth; a further layer of $CX$ gates acting on the Ising chains reduces it to the parity check form of $Z$ and $X$ terms respectively (see Fig. \ref{fig:HamiltonianGS}), and a layer of Hadamards on the $X$ subsystem diagonalises it. The total circuit depth required to get to this form is $3L-1$. Call this gateset $W$. The final form of the circuit can be seen in Fig. \ref{fig:circuitGS}.

Using the result above, to Gibbs sample from the 2 parity check Hamiltonians using local gates, we can replace the unitary $U_n$, which was a product of commuting $CX$ gates to parity check a $n$ qubit computational state, with a weaker unitary $V_n$ that maps 
\begin{align*}
    \ket{x}\ket{+1}& \rightarrow\ket{\sigma(x)}\ket{\pi(x)} \, , \\ \ket{x}\ket{-1}& \rightarrow\ket{\sigma(x)}\ket{-\pi(x)} \, ,
\end{align*}
where $\sigma$ is any bijection of $n$ qubit computational basis states and $\pi$ is the parity; call this a pseudo parity check. We need to then use the fact that the qubits are laid over a 2D lattice, and hence the product operator $\prod_{i} Z_i$ only has diameter $d= \mathcal O(L)$, where the product is taken over all qubits of the $X$ and $Z$ disjoint subsystems respectively. Hence even considering the Lieb-Robinson bounds, we can propose a product of depth $L$ CX gates; any CX gateset that goes through each qubit once and leads towards the ancillae, given by the two empty qubits that would have encoded logical information in the logical state preparation, would do. So we do not even require any additional ancillae. Then after sampling, we have obtained $\sigma(x)$ with the correct statistics of the Ising chain; applying $V^\dagger$ gives us the correct Gibbs state over the $Z, X$ subsystems. Resetting the two ancillae with an $H$ gate and measuring it gives the complete Gibbs state of the parity check form. Further apply $W^\dagger$ to get to the Gibbs state of the toric code in a circuit of $\mathcal O(L)$ depth and $2L^2$ measurements for $L$ even, $2L^2+2$ measurements for $L$ odd, of which $2L^2-2$ can be done simultaneously for both cases. \\

\section{Acknowledgments}
IS is supported by the Cambridge SRIM project and thanks Subhayan Roy Moulik, Owain Salter Fitz-Gibbon and Toshinari C.S. Tong for helpful discussions. 

\bibliography{references}

@article{Gibbs2025,
  author  = {Pablo Páez Velasco and Niclas Schilling and Samuel O. Scalet and Frank Verstraete and Angela Capel},
  title   = {Efficient and simple Gibbs state preparation of the 2D toric code via duality to classical Ising chains},
  journal = {arXiv preprint},
  volume  = {},
  number  = {},
  pages   = {},
  year    = {2025},
  doi     = {}, 
  url     = {https://arxiv.org/abs/2508.00126}
}

@article{Chen2024,
  author  = {Penghua Chen and Bowen Yan and Shawn X. Cui},
  title   = {Quantum circuits for toric code and {X}-cube fracton model},
  journal = {Quantum},
  volume  = {8},
  number  = {},
  pages   = {1276},
  year    = {2024},
  doi     = {10.22331/q-2024-03-13-1276},
  url     = {https://arxiv.org/abs/2210.01682}
}

@article{Bravyi2006,
  author  = {S. Bravyi and M. B. Hastings and F. Verstraete},
  title   = {Lieb-Robinson Bounds and the Generation of Correlations and Topological Quantum Order},
  journal = {Physical Review Letters},
  volume  = {97},
  number  = {5},
  pages   = {050401},
  year    = {2006},
  doi     = {10.1103/PhysRevLett.97.050401},
  url     = {https://arxiv.org/abs/quant-ph/0603121}
}

@article{Gidney2021,
  author  = {Craig Gidney},
  title   = {Stim: a fast stabilizer circuit simulator},
  journal = {arXiv preprint},
  volume  = {},
  number  = {},
  pages   = {},
  year    = {2021},
  doi     = {}, 
  url     = {https://arxiv.org/abs/2103.02202}
}

@article{Claes2025,
  author  = {Jahan Claes},
  title   = {Lower-depth local encoding circuits for the surface code},
  journal = {arXiv preprint},
  volume  = {},
  number  = {},
  pages   = {},
  year    = {2025},
  doi     = {}, 
  url     = {https://arxiv.org/abs/2509.09779}
}

@article{Rouze2024,
  author  = {Cambyse Rouzé and Daniel Stilck França and Álvaro M. Alhambra},
  title   = {Efficient thermalization and universal quantum computing with quantum Gibbs samplers},
  journal = {arXiv preprint},
  volume  = {},
  number  = {},
  pages   = {},
  year    = {2024},
  doi     = {}, 
  url     = {https://arxiv.org/abs/2403.12691}
}

@article{Realization2021,
  author  = {Google Quantum AI},
  journal = {Science},
  volume  = {374},
  number  = {},
  pages   = {},
  year    = {2021},
  doi     = {}, 
  url     = {https://www.science.org/doi/10.1126/science.abi8378}
}

@article{QMH2009,
  title={Quantum Metropolis sampling},
  author={Temme, K. and Osborne, T. J. and Vollbrecht, K. G. and Poulin, D. and Verstraete, F.},
  journal={Nature},
  volume={471},
  number={7336},
  pages={87--90},
  year={2011},
  publisher={Nature Publishing Group},
  doi={10.1038/nature09770}
}

@article{Yung2010,
  author  = {Man-Hong Yung and Alan Aspuru-Guzik},
  title   = {A Quantum-Quantum Metropolis Algorithm},
  journal = {arXiv preprint},
  volume  = {},
  number  = {},
  pages   = {},
  year    = {2010},
  doi     = {}, 
  url     = {https://arxiv.org/abs/1011.1468}
}

@article{Shum2025,
  author  = {Ivan H. C. Shum},
  title   = {Efficient preparation of logical state of the 2D toric code},
  journal = {arXiv preprint},
  volume  = {},
  number  = {},
  pages   = {},
  year    = {2025},
  doi     = {}, 
  url     = {https://arxiv.org/abs/2510.15107}
}

@Book{Levin.2008,
  author     = {Levin, David and Peres, Yuval and Wilmer, Elizabeth},
  publisher  = {American Mathematical Society},
  title      = {Markov Chains and Mixing Times},
  year       = {2008},
  isbn       = {9781470412043},
  month      = dec,
  doi        = {10.1090/mbk/058},
  mrclass    = {60J10 (60-01 60J05 60K35 60K37 68U20 68W20)},
  mrnumber   = {2466937},
  mrreviewer = {Olle\ H\"aggstr\"om},
  pages      = {xviii+371}
}

@InBook{Martinelli.1999,
  author    = {F. Martinelli},
  pages     = {93--191},
  publisher = {Springer Berlin Heidelberg},
  title     = {Lectures on {G}lauber Dynamics for Discrete Spin Models},
  year      = {1999},
  isbn      = {9783540481157},
  booktitle = {Lectures on Probability Theory and Statistics},
  doi       = {10.1007/978-3-540-48115-7_2},
  issn      = {1617-9692},
}

@Book{Brooks.2011,
  author    = {S. Brooks and A. Gelman and G. Jones and X.-L. Meng},
  publisher = {Chapman and Hall/CRC},
  title     = {Handbook of Markov Chain Monte Carlo},
  year      = {2011},
  isbn      = {9780429138508},
  month     = may,
  doi       = {10.1201/b10905},
}

@Article{Rall.2023,
  author    = {Rall, Patrick and Wang, Chunhao and Wocjan, Pawel},
  journal   = {Quantum},
  title     = {Thermal State Preparation via Rounding Promises},
  year      = {2023},
  issn      = {2521-327X},
  month     = oct,
  pages     = {1132},
  volume    = {7},
  doi       = {10.22331/q-2023-10-10-1132},
  publisher = {Verein zur Forderung des Open Access Publizierens in den Quantenwissenschaften},
}

@Article{Wocjan.2023,
  author    = {Wocjan, Pawel and Temme, Kristan},
  journal   = {Communications in Mathematical Physics},
  title     = {Szegedy Walk Unitaries for Quantum Maps},
  year      = {2023},
  issn      = {1432-0916},
  month     = jul,
  number    = {3},
  pages     = {3201--3231},
  volume    = {402},
  doi       = {10.1007/s00220-023-04797-4},
  publisher = {Springer Science and Business Media LLC},
}

@Misc{Jiang.2024,
  author    = {Jiang, Jiaqing and Irani, Sandy},
  title     = {Quantum {Metropolis} Sampling via Weak Measurement},
  year      = {2024},
  eprint    = {2406.16023},
  archivePrefix ={arXiv}
}

@Misc{Gilyen.2024,
  author    = {Gilyén, András and Chen, Chi-Fang and Doriguello, Joao F. and Kastoryano, Michael J.},
  title     = {Quantum generalizations of {G}lauber and {M}etropolis dynamics},
  year      = {2024},
  eprint    = {2405.20322},
  archivePrefix ={arXiv},
}

@article{Kastoryano2016GibbsSamplersCommuting,
  title={Quantum {Gibbs} samplers: {T}he commuting case},
  author={Kastoryano, Michael J and Brandao, Fernando GSL},
  journal={Communications in Mathematical Physics},
  volume={344},
  number={3},
  pages={915--957},
  year={2016},
  publisher={Springer},
  doi={10.1007/s00220-016-2641-8}
}

@article{Ding2024EfficientQuantumGibbs,
  title         = {Efficient Quantum {Gibbs} Samplers with {Kubo}--{Martin}--{Schwinger} Detailed Balance Condition},
  author        = {Z. Ding and B. Li and L. Lin},
  journal       = {Communications in Mathematical Physics},
  year          = {2025},
  volume={406},
  number={3},
  pages={67},
doi = {10.1007/s00220-025-05235-3}
}

@Article{Bardet.2023,
  author    = {I. Bardet and A. Capel and L. Gao and A. Lucia and D. Pérez-García and C. Rouzé},
  journal   = {Physical Review Letters},
  title     = {Rapid Thermalization of Spin Chain Commuting {H}amiltonians},
  year      = {2023},
  issn      = {1079-7114},
  month     = feb,
  number    = {6},
  pages     = {060401},
  volume    = {130},
  doi       = {10.1103/physrevlett.130.060401},
  publisher = {American Physical Society (APS)},
}

@article{Bardet.2024,
  title={Entropy decay for {D}avies semigroups of a one dimensional quantum lattice},
  author={Bardet, Ivan and Capel, {\'A}ngela and Gao, Li and Lucia, Angelo and P{\'e}rez-Garc{\'\i}a, David and Rouz{\'e}, Cambyse},
  journal={Communications in Mathematical Physics},
  volume={405},
  number={2},
  pages={42},
  year={2024},
  publisher={Springer},
  doi = {10.1007/s00220-023-04869-5},
}

@article{Capel2024Gibbs,
      title={Quasi-optimal sampling from {G}ibbs states via non-commutative optimal transport metrics}, 
      author={A. Capel and P. Gondolf and J. Kochanowski and C. Rouzé},
      journal   = {Annales Henri Poincaré},
    volume = {},
    number = {},
  year      = {2025},
doi = {10.1007/s00023-025-01637-0},
}

@article{Capel2024RapidThermalizationDissipative,
  title     = {Rapid Thermalization of Dissipative Many-Body Dynamics of Commuting {H}amiltonians},
  author    = {J. Kochanowski and A. M. Alhambra and A. Capel and C. Rouzé},
  journal   = {Communications in Mathematical Physics},
volume = {406},
number = {176},
  year      = {2025},
doi = {article/10.1007/s00220-025-05353-y},
}

@Misc{Bakshi.2024,
  author    = {A. Bakshi and A. Liu and A. Moitra and E. Tang},
  title     = {High-Temperature {G}ibbs States are Unentangled and Efficiently Preparable},
  year      = {2024},
  eprint    = {2403.16850},
  archivePrefix = {arXiv},
}

@misc{hwang2024gibbsstatepreparationcommuting,
      title={{G}ibbs state preparation for commuting {H}amiltonian: {M}apping to classical {Gibbs} sampling}, 
      author={Y. Hwang and J. Jiang},
      year={2024},
      eprint={2410.04909},
      archivePrefix={arXiv}
}

@misc{schmidhuber2025hamiltoniandecodedquantuminterferometry,
      title={Hamiltonian Decoded Quantum Interferometry}, 
      author={Alexander Schmidhuber and Jonathan Z. Lu and Noah Shutty and Stephen Jordan and Alexander Poremba and Yihui Quek},
      year={2025},
      eprint={2510.07913},
      archivePrefix={arXiv},
      primaryClass={quant-ph},
      url={https://arxiv.org/abs/2510.07913}, 
}

\vspace{1cm}

\appendix

\section{Classical Gibbs sampling algorithms}\label{appendix}

We provide explicit algorithms to Gibbs sample from:
\begin{itemize}
    \item The classical magnetic field:
    \begin{equation*}
        H_1=-\sum_{i=1}^n Z_i \, .
    \end{equation*}
    \item The Ising chain with magnetic field on one end:
    \begin{equation*}
        H_2=-\sum_{i=1}^{n-1} Z_i Z_{i+1} -Z_n \, .
    \end{equation*}
    \item The Ising loop:
    \begin{equation*}
        H_3=\sum_{i=1}^{n-1} Z_i Z_{i+1} -Z_n Z_1 \, .
    \end{equation*}
    \item The parity check Hamiltonian:
    \begin{equation*}
        H_4=-\sum_{i=1}^{n-1} Z_i-\Pi_{i=1}^{n-1} Z_i \, ,
    \end{equation*}
     called this since the last term penalises odd parity states, and is defined over $n$ qubits. 
\end{itemize}
The Gibbs state of $H_1$ over $M$ qubits where $M>n$, is given by a cartesian product of $M$ Bernoulli variables, $n$ of which have parameter $p=\dfrac{1}{1+e^{2\beta}}$, and $M-n$ of which have parameter $p=1/2$. This can be prepared by a quantum circuit of depth 1 and a simultaneous measurement in the computational basis on all qubits, by rotating each of the qubits with a rotation $e^{-i\theta Y}$ that gives the Boltzmann weights: $\theta=\tan^{-1}e^{-\beta}$ for the first $n$ qubits where the Hamiltonian acts non-trivially and $\theta=\pi/4$ else. 

The Ising chain with a magnetic field on one end $H_2$ is then unitarily equivalent to the Hamiltonian above: conjugating $H_2$ with $CX(j+1, j)$ for $j=1,2,.,n-1$ in this order gives $H_1$, and hence to obtain samples from the Gibbs state of $H_2$, we sample from $H_1$ and then apply $XOR(j+1, j)$ in order $j=n-1,n-2,..,1$. 

Similarly, we can Gibbs sample from 
\begin{equation*}
    H_0=-\sum_{i=1}^{n-2} Z_i Z_{i+1} -Z_{n-1}-Z_1 \, ,
\end{equation*}
an Ising chain with a magnetic field on both ends, defined over $n$ qubits, as seen in \cite{Gibbs2025}, while noting the two Hamiltonians $H_3, H_4$ are unitarily equivalent; the Gibbs state of $H_4$ can be reached from that of $H_0$ by applying $\prod_{i=1}^{n-1}XOR(n, i)$; $H_3$ is unitarily equivalent to $H_4$ by conjugating $XOR(j+1, j)$ for $j=1,2,..,n-1$ in order, and hence the Gibbs state is achieved by apply these layers of $XOR$ in reverse order, to the sampling result of $H_4$.

\end{document}